\documentclass[twocolumn]{aastex62}

\usepackage{amsmath}
\usepackage{bm}
\usepackage{color}
\usepackage{ulem}
\accepted{July 31, 2020}

\shorttitle{Nonlinear Alfv\'en Waves in the Solar Chromosphere}
\shortauthors{Sakaue \& Shibata}

\begin{document}

\title{Energy Transfer by Nonlinear Alfv\'en Waves in the Solar Chromosphere, and Its Effect on Spicule Dynamics, Coronal Heating, and Solar Wind Acceleration}

\correspondingauthor{Takahito Sakaue}
\email{sakaue@kwasan.kyoto-u.ac.jp}


\author{Takahito Sakaue}
\affiliation{Astronomical Observatory, Kyoto University, Japan}

\author{Kazunari Shibata}
\affiliation{Astronomical Observatory, Kyoto University, Japan}






\begin{abstract}
  Alfv\'en waves are responsible for the transfer of magnetic energy in the magnetized plasma. They are involved in heating solar atmosphere and driving solar wind through various nonlinear processes. Since the magnetic field configurations directly affect the nonlinearity of Alfv\'en waves, it is important to investigate how they relate to the solar atmosphere and wind structure through the nonlinear propagation of Alfv\'en waves.
  In this study, we carried out the one-dimensional magnetohydrodynamic simulations to realize the above relation.
  The results show that when the nonlinearity of Alfv\'en waves in the chromosphere exceeds a critical value, the dynamics of the solar chromosphere (e.g., spicule) and the mass loss rate of solar wind tend to be independent of the energy input from the photosphere. In a situation where the Alfv\'en waves are highly nonlinear, the strong shear torsional flow generated in the chromosphere ``fractures'' the magnetic flux tube. This corresponds to the formation of chromospheric intermediate shocks, which limit the transmission of the Poynting flux into the corona by Alfv\'en waves and also inhibits the propagation of chromospheric slow shock.

\end{abstract}

\keywords{Sun: chromosphere, solar wind, magnetohydrodynamics (MHD)}


\section{INTRODUCTION} \label{sec:intro}

The solar atmosphere consists of magnetized plasma with various thermal properties. One MK corona is characterized by tenuous, fully ionized, and low $\beta$ plasma. It is the envelope of a cool ($\sim10^4$ K), dense, and partially ionized chromosphere. The coronal and chromospheric heating problems arise from the question regarding the manner of steadily supplying and depositing the energy to maintain such a thermal structure of solar atmosphere. These problems are directly related to the physical mechanism for the solar wind acceleration.

The nonlinear propagation of Alfv\'en waves are one of the promising physical mechanisms in solving this problem. That is because this incompressible wave is responsible for the transfer of magnetic energy in the magnetized plasma and is involved in the energy conversion to the kinetic or thermal energy of the background media through the nonlinear processes.
Numerous theoretical studies have developed the scenario relating the Alfv\'en waves to atmospheric heating \citep{1947MNRAS.107..211A,1961ApJ...134..347O,1968ApJ...153..371C,1983A&A...117..220H}, the solar wind acceleration \citep{1976ApJ...210..498B,1980JGR....85.1311H}, and the spicule dynamics \citep{1982SoPh...75...35H,1999ApJ...514..493K}.
These ideas have been examined using the spaceborne observations that confirmed the ubiquitous existence of the Alfv\'en waves from the chromosphere \citep{2007Sci...318.1574D,2011ApJ...736L..24O}, corona \citep{2007Sci...318.1580C,2009A&A...501L..15B,2013ApJ...776...78H} to the interplanetary space \citep{1971JGR....76.3534B,1982JGR....87.3617B,2001JGR...10610659B}.

Recent magnetohydrodynamics (MHD) simulations enable a more seamless description about the relationship between the Alfv\'en wave propagation and the dynamics of solar atmosphere and wind. Because of the inhomogeneous, time-dependent, and stratified solar atmosphere, the Alfv\'en wave propagation can be affected by various physical mechanisms in each layer of the solar atmosphere. \citet{2012ApJ...749....8M} and \citet{2014MNRAS.440..971M} carried out the 2.5-dimensional simulation and showed the self-consistent transition of heating mechanisms from shock heating to the incompressible processes across the transition layer. On the basis of their 3D simulation, \citet{2019ApJ...880L...2S} confirmed that the density fluctuation caused by the parametric decay instability \citep{1978ApJ...219..700G,1978ApJ...224.1013D,1986JGR....91.4171T} is essential in exciting Alfvén wave turbulence in the solar wind.

Aside from the abovementioned multi-dimensional models, one-dimensional (1D) simulations are still helpful, particularly in investigating the diversity or universality of solar and stellar atmosphere and wind. They have contributed to understanding how the Alfv\'en waves are involved with the spicule \citep{1982SoPh...75...35H,2010ApJ...710.1857M}, the solar and stellar wind \citep{2005ApJ...632L..49S,2007ApJ...659.1592S,2018PASJ...70...34S,2019ApJ...879...77Y}, and the coronal loop \citep{2004ApJ...601L.107M,2010ApJ...712..494A,2019ApJ...885..164W}.
Despite these extensive works, there have been few studies focused on the chromospheric magnetic field environment in terms of the influence on the solar atmosphere and wind. The magnetic field in the solar atmosphere is highly inhomogeneous and variable with time. Thus, it directly affects the profile of the Alfv\'en speed with respect to the height, which determines the reflection efficiency of the Alfv\'en waves \citep{1990ApJ...350..309A,1993A&A...270..304V} and induces Alfv\'en resonance \citep{1978SoPh...56..305H,2010ApJ...710.1857M}. The expanding magnetic flux tube in lower atmosphere, additionally, is related to the rapid evolution of the Alfv\'en wave amplitude. That leads to the dissipation of the Alfv\'en waves through direct steepening \citep{1982SoPh...75...35H} or nonlinear mode coupling \citep{1992ApJ...389..731H,1999ApJ...514..493K,2020ApJ...891..110W}. The coronal heating and solar wind acceleration are sustained with slight transmission of the Alfv\'en waves from the chromosphere. Therefore, it is worthwhile to examine how robustly Alfv\'en waves can transport the magnetic energy across the chromosphere even in the different magnetic field configurations in lower atmosphere.

In this study, we performed time-dependent 1D MHD simulations similar to \citet{1999ApJ...514..493K} or \citet{2005ApJ...632L..49S}. Unlike them, we pay particular attention to the dependence of the spicule dynamics, coronal heating and solar wind acceleration on the magnetic field configuration in the lower atmosphere.


\section{NUMERICAL SETTING}
\subsection{Basic Equations}
\label{sec:basic_equations}

We used 1D magnetohydrodynamic equations based on the axial symmetry assumption of the magnetic flux tube.
The surface of the axisymmetric flux tube is defined by the poloidal and toroidal axes which are noted in this study with $x$ and $\phi$. The basic equations in CGS unit are written as follows:

The mass conservation law is presented by:
\begin{equation}
  {\partial\rho\over\partial t}+{1\over A}{\partial\over\partial x}(\rho v_xA)=0\label{eq:mass}
\end{equation}
where $\rho$, $v_x$ and $A$ are the mass density, poloidal component of velocity, and cross section of the flux tube, respectively.

The energy conservation law is presented by:
\begin{align}
  {\partial\over\partial t}&\left({p\over\gamma-1}+{1\over2}\rho v^2+{B^2\over8\pi}\right)\nonumber\\
  +&{1\over A}{\partial\over\partial x}\left[A\left\{\left({\gamma p\over\gamma-1}+{\rho v^2\over2}+{B_\phi^2\over4\pi}\right)v_x-{B_x\over4\pi}(B_\phi v_\phi)\right\}\right]\nonumber\\
  &=\rho v_x{\partial\over\partial x}\left({GM_\odot\over r}\right)-{1\over A}{\partial\over\partial x}(AF_c)-Q_{\mbox{\scriptsize rad}}\label{eq:ene}
\end{align}
where $p$, $B_x$, $B_\phi$, $v_\phi$ and $\gamma$ are the gas pressure, poloidal and toroidal components of magnetic field, toroidal component of velocity, and the specific heat ratio which is set to 5/3, respectively. $v^2=v_x^2+v_\phi^2$ and $B^2=B_x^2+B_\phi^2$. $G$ and $M_\odot$ are gravitational constant and the solar mass. $r$ is the distance from the sun center. $F_c$ and $Q_{\rm rad}$ represent the heat conduction flux and radiative cooling term, respectively, as described in section \ref{sec:Heat Conduction and Radiative Cooling}.\par
The poloidal component of the equation of motion is presented by:
\begin{align}
  {\partial(\rho v_x)\over\partial t}+&{\partial p\over\partial x}+{1\over A}{\partial\over\partial x}\left\{\left(\rho v_x^2+{B_\phi^2\over8\pi}\right)A\right\}\nonumber\\
  &-\rho v_\phi^2{\partial\ln\sqrt{A}\over\partial x}-\rho{\partial\over\partial x}\left({GM_\odot\over r}\right)=0\label{eq:mom_pol}
\end{align}

The toroidal component of the equation of motion is presented by:
\begin{equation}
  {\partial(\rho v_\phi)\over\partial t}+{1\over A\sqrt{A}}
  {\partial\over\partial x}\left\{A\sqrt{A}\left(\rho v_xv_\phi-{B_xB_\phi\over4\pi}\right)\right\}=0\label{eq:mom_tor}
\end{equation}

The toroidal component of the induction equation is presented by:
\begin{equation}
  {\partial B_\phi\over\partial t}+{1\over \sqrt{A}}{\partial\over\partial x}\Big(\sqrt{A}(v_xB_\phi-v_\phi B_x)\Big)=0\label{eq:mag_tor}
\end{equation}

The poloidal magnetic flux conservation is presented by:
\begin{equation}
  B_xA=\mbox{const.}\label{eq:mag_pol}
\end{equation}

Finally, we note that the poloidal axis $x$ is not always parallel to the radial axis $r$. They are related to each other as follows:
\begin{equation}
  {dx\over dr}=\sqrt{1+\left({d\sqrt{A}\over dr}\right)^2}
\end{equation}

\subsection{Magnetic Flux Tube Model}
\label{sec:magnetic_flux_tube_model}

The assumption of the background magnetic field is described here in detail. The cross section of the flux tube $A$ is related to $r$ through the filling factor $f$ as $A(r)=4\pi r^2f(r)$. $f$ determines the geometry of the flux tube. We consider the axisymmetric magnetic flux tube from the photosphere to the interplanetary space. The outer boundary of our simulation is set to 0.5 AU.
In the lower atmosphere, the magnetic flux tube expands exponentially such that the magnetic pressure inside the flux tube balances out with the ambient plasma gas pressure which decreases with the scale height $H_{\rm ph}=R_gT_{\rm eff}/({\mu_{\rm ph}} g_\odot)$. Here, $R_g=8.31\times10^7$ erg K$^{-1}$ mol$^{-1}$ is the gas constant, $T_{\rm eff}=5770$ K, $\mu_{\rm ph}=1.3$ is the mean molecular weight on the photosphere, and $g_\odot$ is the gravitational acceleration on the solar surface. The filling factor $f$ in this layer is expected to be $f_{\rm atm}(r)=f_{\rm ph}\exp\{r_\odot/(2H_{\rm ph})(1-r_\odot/r)\}$, where $f_{\rm ph}$ is the coverage of the open magnetic flux tube on the photosphere. By using $f_{\rm atm}$, $B_x=r_\odot^2f_{\rm ph}B_{\rm ph}/(r^2f_{\rm atm})$ satisfies the condition that $B_x^2/(8\pi)=p_{\rm atm}$, where $p_{\rm atm}$ is the solution of the hydrostatic equilibrium. In the lower atmosphere where $r=r_\odot+h$ ($h\ll r_\odot$), we obtain $f_{\rm atm}(h)=f_{\rm ph}e^{h/(2H_{\rm ph})}$. This exponential expansion of flux tube is assumed to stop at some height where it merges with the neighboring flux tube. Above this height (i.e., the merging height $H_m$), the magnetic pressure dominates the gas pressure and the flux tube extends vertically. The poloidal magnetic field strength in this layer is assumed to be almost constant around $\overline{B}=B_{\rm ph}f_{\rm ph}/f_{\rm atm}(H_m)=B_{\rm ph}e^{-H_m/(2H_{\rm ph})}$ through the upper chromosphere and coronal base. Thus, $\overline{B}=B_{\rm ph}e^{-H_m/(2H_{\rm ph})}$ roughly represents the area-averaged magnetic field strength in the coronal hole from which the solar wind emanates. It should be noted that the various flux tube models in the lower atmosphere have been considered, for example, by \citet{2003ApJ...585.1138H,2005ApJ...631.1270H} and \citet{2005ApJS..156..265C}. The different magnetic field geometries would lead to different results. Their significance should be tested in future studies as long as we rely on the 1D simulation.

\begin{figure*}
  \begin{center}
    \epsscale{1.2}  
    \plotone{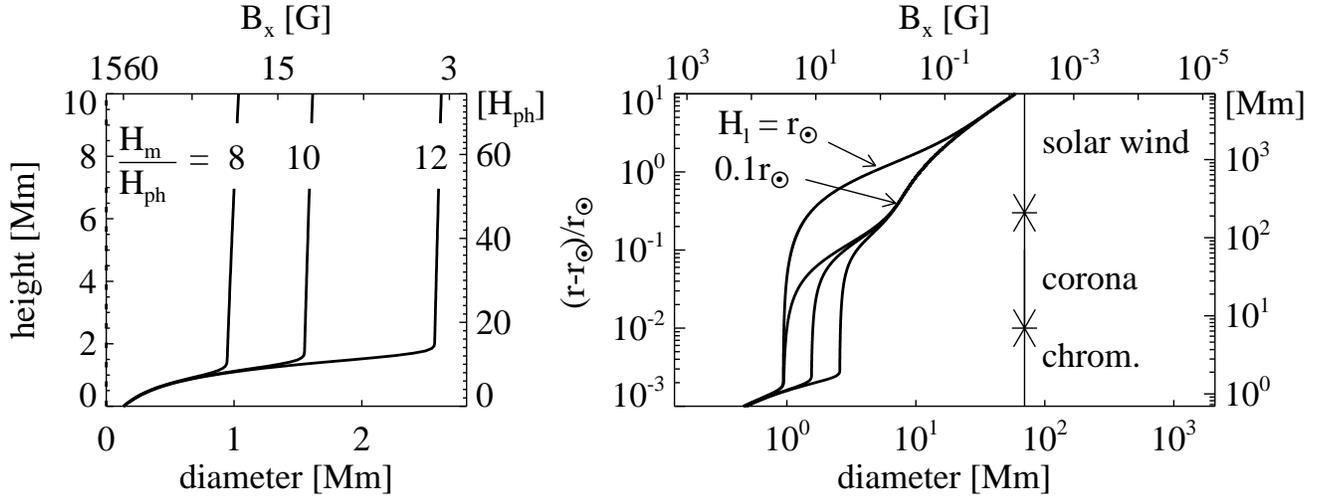}
    \caption{Poloidal magnetic field configurations characterized with the free parameters ($H_m$, $H_l$), where $H_m$, $H_l$ and $H_{\rm ph}$ are the merging height, loop height, and the pressure scale height on the photosphere, respectively.  Left and right panels show it in the lower and outer atmosphere.}
    \label{fig:fig01.eps}
  \end{center}
\end{figure*}

The flux tube is assumed to expand super radially again in the extended corona such that the interplanetary space is filled with the open flux tube. We characterize this expansion with the coronal loop height $H_l$. The functional form of the filling factor in this layer $f_{\rm wind}(r)$ is suggested by \citet{1976SoPh...49...43K}. Based on these considerations, the profile of the filling factor $f(r)$ is determined as follows:
\begin{equation}
  f_{\rm atm}(r)=f_m\tanh\left[{f_{\rm ph}\over f_m}\exp\left\{{r_\odot\over 2H_{\rm ph}}\left(1-{r_\odot\over r}\right)\right\}\right]
  \label{eq:f_eq1}
\end{equation}
where $f_m=f_{\rm ph}e^{H_m/(2H_{\rm ph})}$
\begin{equation}
  f_{\rm wind}(r)={e^{(r-r_\odot-H_l)/\sigma_l}+f_m-(1-f_m)e^{-(H_l/\sigma_l)}
    \over e^{(r-r_\odot-H_l)/\sigma_l}+1}
  \label{eq:f_eq2}
\end{equation}
\begin{align}
  \hat{f}(r)=f_{\rm atm}(r)+{1\over2}&\Big(\max[f_{\rm wind}(r),f_m]-f_{\rm atm}(r)\Big)\nonumber\\
  &\times\left\{1+\tanh\left({r-r_\odot-H_l\over H_l}\right)\right\}
  \label{eq:f_eq3}
\end{align}
\begin{equation}
  f(r)=f_{\rm ph}+(1-f_{\rm ph}){\hat{f}(r)-\hat{f}(r_\odot)\over1-\hat{f}(r_\odot)}
  \label{eq:f_eq4}
\end{equation}
The key parameters of $f(r)$ are $f_{\rm ph}$, $H_m$, and $H_l$. $\sigma_l$ in Eq. \ref{eq:f_eq2} is set to $H_l$. The manner by which the properties of solar and stellar wind depend on $f_{\rm ph}$ has already been well-investigated in previous studies \citep{2006ApJ...640L..75S,2013PASJ...65...98S}. Thereafter, we use the fixed value of 1/1600 for $f_{\rm ph}$ by referring to \cite{2013PASJ...65...98S}. Note that, when $f_{\rm ph}=1/1600$, the magnetic field strength at $r=$1 AU is 2.1nT, which is within the typical observed value \citep{2000GeoRL..27..505W}. The configuration of the magnetic flux tube with $f_{\rm ph}$=1/1600 is depicted in Figure \ref{fig:fig01.eps}. As shown in this figure, the merging height $H_m$ is the parameter defining the magnetic field strength $\overline{B}$ from the chromosphere up to the lower corona. The higher merging height corresponds to a weaker magnetic field $\overline{B}$, and, in particular, $H_m/H_{\rm ph}=8,12$ are used in this study. It should be noted that $H_m/H_{\rm ph}=8,10,12$ correspond to $\overline{B}=29,11,4$ G, respectively. These magnetic field strengths are comparable to the typical value for the area-averaged magnetic field strength in the coronal hole (3 to 36 G near the solar activity maximum and 1 to 7 G close to the minimum, according to \citet{1982SoPh...79..149H}; see also review by \citet{2004SoPh..225..227W}).
By adopting a higher coronal loop height $H_l$, the magnetic field strength in the upper corona can be larger (Figure \ref{fig:fig01.eps}), but $H_l/r_\odot$ is fixed at 0.1.

\subsection{Heat Conduction and Radiative Cooling}
\label{sec:Heat Conduction and Radiative Cooling}
The equation of state is $p=\rho R_gT/[\mu_{\rm ph}(1-\chi(T)/2)]$, where $\chi(T)$ is the ionization degree as a function of temperature which is calculated by referring to \citet{2012A&A...539A..39C}. The radiative cooling $Q_{\rm rad}$ is given by the empirical formulae, which is composed of three distinct terms, i.e., the photospheric radiation $Q_{\rm ph}$, chromospheric radiation $Q_{\rm ch}$, and coronal radiation $Q_{\rm cr}$.
\begin{equation}
  Q_{\rm rad}=(1-\xi_1)(1-\xi_2)Q_{\rm ph}+\xi_1(1-\xi_2)Q_{\rm ch}+\xi_2Q_{\rm cr}
  \label{eq:total_radiation}
\end{equation}
where $\xi_1$ and $\xi_2$ are assumed to be as follows:
\begin{equation}
  \xi_1={1\over2}\left[1+\tanh\left({r-r_\odot\over H_{\rm ph}}-3\right)\right]
\end{equation}
\begin{equation}
  \xi_2=\exp\left(-4\times10^{-20}\int_\infty^rn_{\rm HI}dr'\right)
\end{equation}
$n_{\rm HI}=(1-\chi(T))\rho/m_p$ is the neutral hydrogen density, where $m_p$ is the proton mass.
Each term in Eq. \ref{eq:total_radiation} is defined as follows
\begin{equation}
  Q_{\rm ph}=4\rho\kappa_R\sigma_{\rm SB} T^4\max\left({T^4\over T_{\rm ref}^4}-1,-e^{-(r-r_\odot)^2/H_{\rm ph}^2}\right)
  \label{eq:eq_q_ph}
\end{equation}
\begin{equation}
  \mbox{where}\ \ T_{\rm ref}=T_{\rm eff}\left({3\over4}\rho\kappa_RH_{\rm ph}+{1\over2}\right)^{1/4}
\end{equation}
\begin{equation}
  Q_{\rm ch}=4.9\times10^9{\ \small[\mbox{erg g$^{-1}$ s$^{-1}$}]\ }\rho,\ \ \ Q_{\rm cr}=\chi(T)n^2\Lambda(T) \label{eq:eq_q_cr}
\end{equation}
$\kappa_R=0.2$ cm$^2$ g$^{-1}$ pertains to the Rosseland opacity on the photosphere. $\sigma_{SB}$ is the Stefan–Boltzmann constant. $n$ is the number density of neutral or ionized hydrogen; i.e., $n=\rho/m_p$. $\Lambda(T)$ is the radiative loss function for the optically thin plasma. $Q_{\rm ch}$ and $\Lambda(T)$ are the same function as used in \citet{1997ApJ...489..426H}, which are always positive. $Q_{\rm ph}$ in Eq. \ref{eq:eq_q_ph} is allowed to be negative where $e^{-(r-r_\odot)^2/H_{\rm ph}^2}\sim1$, which represents the radiative heating.

The heat conductive flux is presented by
\begin{equation}
  F_{\rm c}=-\kappa(T){\partial T\over\partial x}
\end{equation}
where $\kappa(T)$ is the heat conductivity as a function of the temperature. That is composed of the collisional and collisionless terms:
\begin{equation}
  \kappa(T)=q\kappa_{\rm coll}+(1-q)\kappa_{\rm sat}
\end{equation}
where $q=\max(0,\min(1,1-0.5\kappa_{\rm coll}/\kappa_{\rm sat}))$. $\kappa_{\rm coll}(T)$ is adopted from \citet{1980SoPh...68..351N}, which agrees with the Spitzer-H\"arm heat conductivity \citep{1953PhRv...89..977S} $\kappa_0 T^{5/2}$ ($\kappa_0=10^{-6}$ in CGS unit) when $T>10^6$ K. $\kappa_{\rm sat}$ is presented by
\begin{equation}
  \kappa_{\rm sat}={3\over2}pv_{e,\rm thr}{r\over T}
\end{equation}
where $v_{e,\rm thr}$ is the thermal speed of the electron. $\kappa_{\rm sat}$ represents the saturation of heat flux caused by the collisionless effect \citep{1964ApJ...139...93P,2013ApJ...769L..22B}.  The above expression of $\kappa_{\rm sat}$ means that the transition of heat conductivity from $\kappa_{\rm coll}$ to $\kappa_{\rm sat}$ occurs around $r\sim\lambda_{e,\rm mfp}$ ($\lambda_{e,\rm mfp}$ is the electron mean free path) and that the heat flux is limited to ${3\over2}\alpha pv_{e,\rm thr}$ in the distance where $T\sim r^{-\alpha}$ ($\alpha=0.2-0.4$ for winds faster than $500$ km s$^{-1}$; \citet{1989JGR....94.6893M}).
Based on the foregoing heat conductivity, heat conduction is solved by the super-time-stepping method \citep{2012MNRAS.422.2102M,2014JCoPh.257..594M}.

\subsection{Initial and Boundary Condition}
\label{sec:Initial_and_Boundary_Condition}

We set the static atmosphere with a temperature of $10^4$ K as the initial state. The temperature on the bottom boundary is promptly cooled down to $T_{\rm eff}=$5770 K after the initiation of the simulation. The mass density and poloidal magnetic field strength on the photosphere are $\rho_{\rm ph}=2.5\times10^{-7}$ g cm$^{-3}$ and $B_{\rm ph}=$1560 G, respectively. To excite the outward propagating Alfv\'en wave, the toroidal velocity $v_\phi$ on the bottom boundary is oscillated artificially, which represents the convective motion on the solar photosphere. We consider it as a frequency-dependent fluctuation with the following power spectrum.
\begin{align}
  v_{\rm conv}^2\propto\int^{\nu_{\rm max}}_{\nu_{\rm min}}\nu^{-1}d\nu
\end{align}
where $v_{\rm conv}$ is the free parameter corresponding to the amplitude of the convective velocity. $\nu_{\rm min}^{-1}$ and $\nu_{\rm max}^{-1}$ are 30 min and 20 s, respectively. The phase offsets of fluctuation are randomly assigned. The amplitude of fluctuation $v_{\rm conv}$ is the subject of survey in this study, e.g., $v_{\rm conv}/c_{s\rm ph}=$0.07, 0.14, 0.21, 0.42, 0.85 ($c_{s\rm ph}=\sqrt{\gamma R_gT_{\rm eff}/\mu_{\rm ph}}=7.8$ km s$^{-1}$ is the adiabatic sound speed on the photosphere). This parameter range includes the typical velocity of horizontal convective motion; 1.1 km s$^{-1}$ \citep{2010ApJ...716L..19M}.

To excite the purely outward Alfv\'en waves on the bottom boundary, the toroidal magnetic field $B_\phi$ is determined by $B_\phi=-\sqrt{4\pi\rho} v_\phi$. This means that the Els\"asser variables (i.e., $z_{\rm out}=v_\phi-B_\phi/\sqrt{4\pi\rho}$, and $z_{\rm in}=v_\phi+B_\phi/\sqrt{4\pi\rho}$) on the bottom boundary satisfy the conditions, i.e., $z_{\rm out}=2v_\phi$ and $z_{\rm in}=0$. The longitudinal velocity component $v_x$ on the bottom boundary is also given as the fluctuation with the amplitude $v_{\rm conv}$, the power spectrum similar to that of the foregoing, and the randomly assigned phase offsets. We performed a few simulations with $v_x=0$ on the bottom boundary. We were able to confirm that $v_x\neq0$ on the photosphere does not have any influence on the solar wind structure, but the spicule height can depend on it.

The upper boundary is treated as the free boundary. 
19200 grids are placed nonuniformly in between. The numerical scheme is based on the HLLD Riemann solver \citep{2005JCoPh.208..315M} with the second-order MUSCL interpolation and the third-order TVD Runge–-Kutta method \citep{1988JCoPh..77..439S}.

\section{RESULTS}
\label{sec:sec_results}







\subsection{Solar Wind Profiles}
\label{sec:solar_wind_profiles}
\begin{figure*}
  \begin{center}
    \epsscale{.8}  
    \plotone{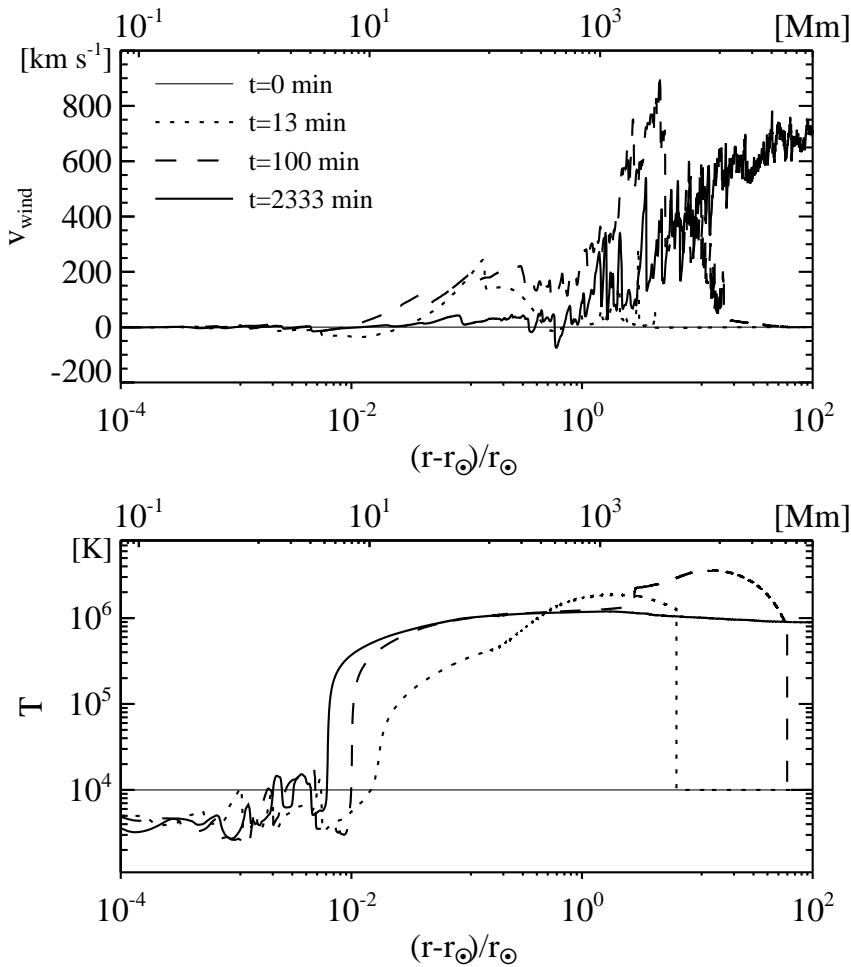}
    \caption{
      The temporal variations of the solar wind velocity (upper) and temperature (lower) given that the simulation starts in the case of $\overline{B}=29$ G and $v_{\rm conv}/c_{s\rm ph}=0.21$.
    }
    \label{fig:fig02.eps}
  \end{center}
\end{figure*}
\begin{figure}
  \begin{center}
    \epsscale{1.2}  
    \plotone{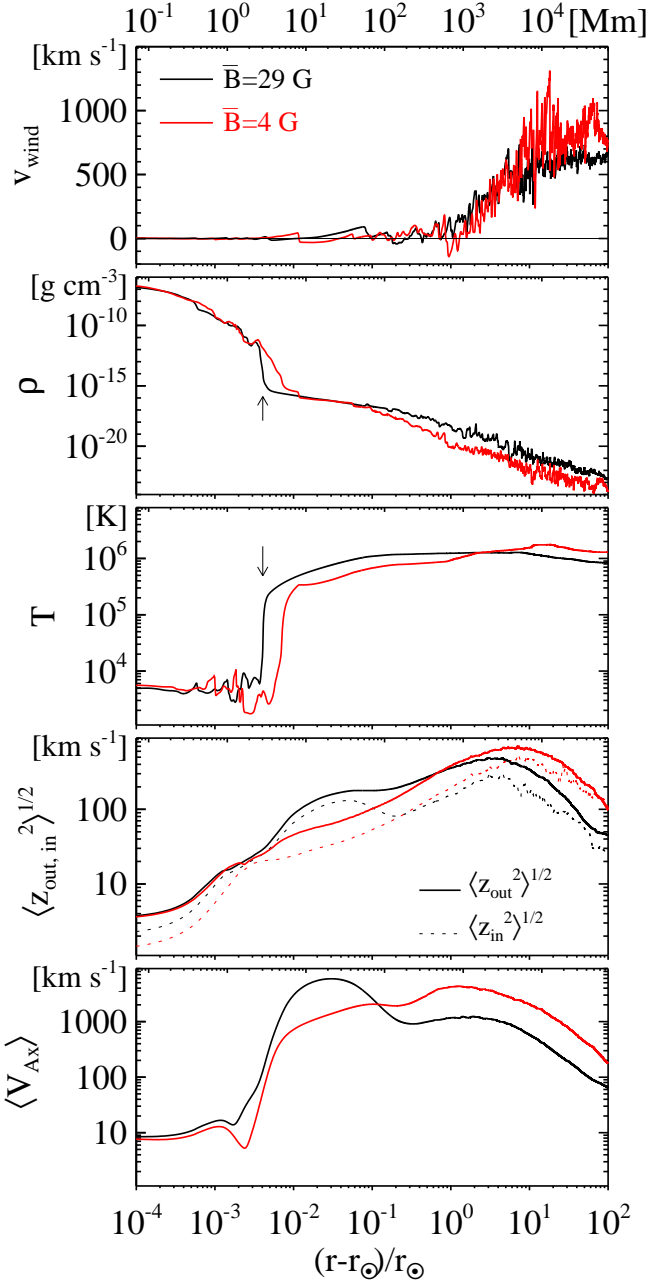}
    \caption{
      The snapshots of solar wind velocity, mass density, temperature profiles, and temporally averaged profiles of Alfv\'en wave amplitude and Alfv\'en speed in the solar wind. The black and red lines represent the profiles in the cases of $\overline{B}=29$ and 4 G, respectively. $v_{\rm conv}/c_{s\rm ph}=0.21$. The corresponding times of the presented snapshots is $t=62$ hrs in the $\overline{B}=29$ G case and $t=45$ hrs in the $\overline{B}=4$ G case. The arrows in the second and third panels indicate the top of chromosphere ($T=4\times10^4$ K, $\rho=10^{-14}$ g cm$^{-3}$).
    }
    \label{fig:fig03.eps}
  \end{center}
\end{figure}
\noindent
\begin{figure}
  \begin{center}
    \epsscale{1.2}  
    \plotone{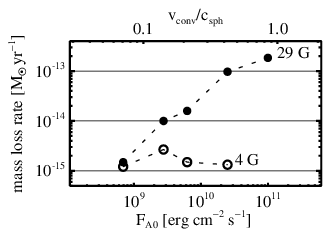}
    \caption{
      The mass loss rates of solar wind as a function of the energy input from the photosphere ($F_{A0}$). The filled and open symbols correspond to the simulation results with the $\overline{B}=29$ and 4 G.
    }
    \label{fig:fig04.eps}
  \end{center}
\end{figure}
After several tens of hours, the solar wind in the simulation box reaches the quasi-steady state with the numerous wave signatures (Figure \ref{fig:fig02.eps}). Figure \ref{fig:fig03.eps} shows the simulation results, including the snapshots of solar wind velocity, mass density, temperature profiles, and temporally averaged profiles of Alfv\'en wave amplitude and Alfv\'en speed in the solar wind. The black and red lines in each figure correspond to the results in the cases of $\overline{B}=29$ and 4 G, respectively.

The top panel of Figure \ref{fig:fig03.eps} shows that the solar wind in $\overline{B}=4$ G case is found to be faster than that in $\overline{B}=29$ G case. Alfv\'en speed at the coronal base is much higher in $\overline{B}=29$ G than in $\overline{B}=4$ G. In the outer space above the coronal loop height, where the magnetic field strengths in both cases are the same, Alfv\'en speed in $\overline{B}=4$ G is larger than in $\overline{B}=29$ G, clearly indicating the denser wind in $\overline{B}=29$ G. With regard to the higher Alfv\'en speed at the coronal base in $\overline{B}=29$ G, Alfv\'en speed steeply declines above the coronal loop height due to the largely expanding magnetic flux tube. This induces the strong interference between the outward and inward Alfv\'en waves, resulting in the humps of the Alfv\'en wave amplitude profiles below $0.1r_\odot$.

The most significant discrepancy between the solar winds in the different merging heights is found in the wind's mass loss rate. Figure \ref{fig:fig04.eps} shows the mass loss rates as a function of the energy input from the photosphere ($F_{A0}=\rho_{\rm ph}v_{\rm conv}^2V_{A\rm ph}$). The filled and open circles show the results for $\overline{B}=29$ and 4 G, respectively. While the wind's mass loss rate monotonically increases with a larger energy input from the photosphere in the case of $\overline{B}=29$ G, that in $\overline{B}=4$ G is almost independent of the energy input. The mass loss rate in $\overline{B}=4$ G is limited to $\sim10^{-15}$ $M_\odot$ yr $^{-1}$ even in the largest energy input case of $v_{\rm conv}/c_{s\rm ph}=0.42$, which is two orders of magnitude smaller than that in the $\overline{B}=29$ G case.

\subsection{Spicule Dynamics}
\label{sec:spicule_dynamics}
\begin{figure*}
  \begin{center}
    \epsscale{1}
    \plotone{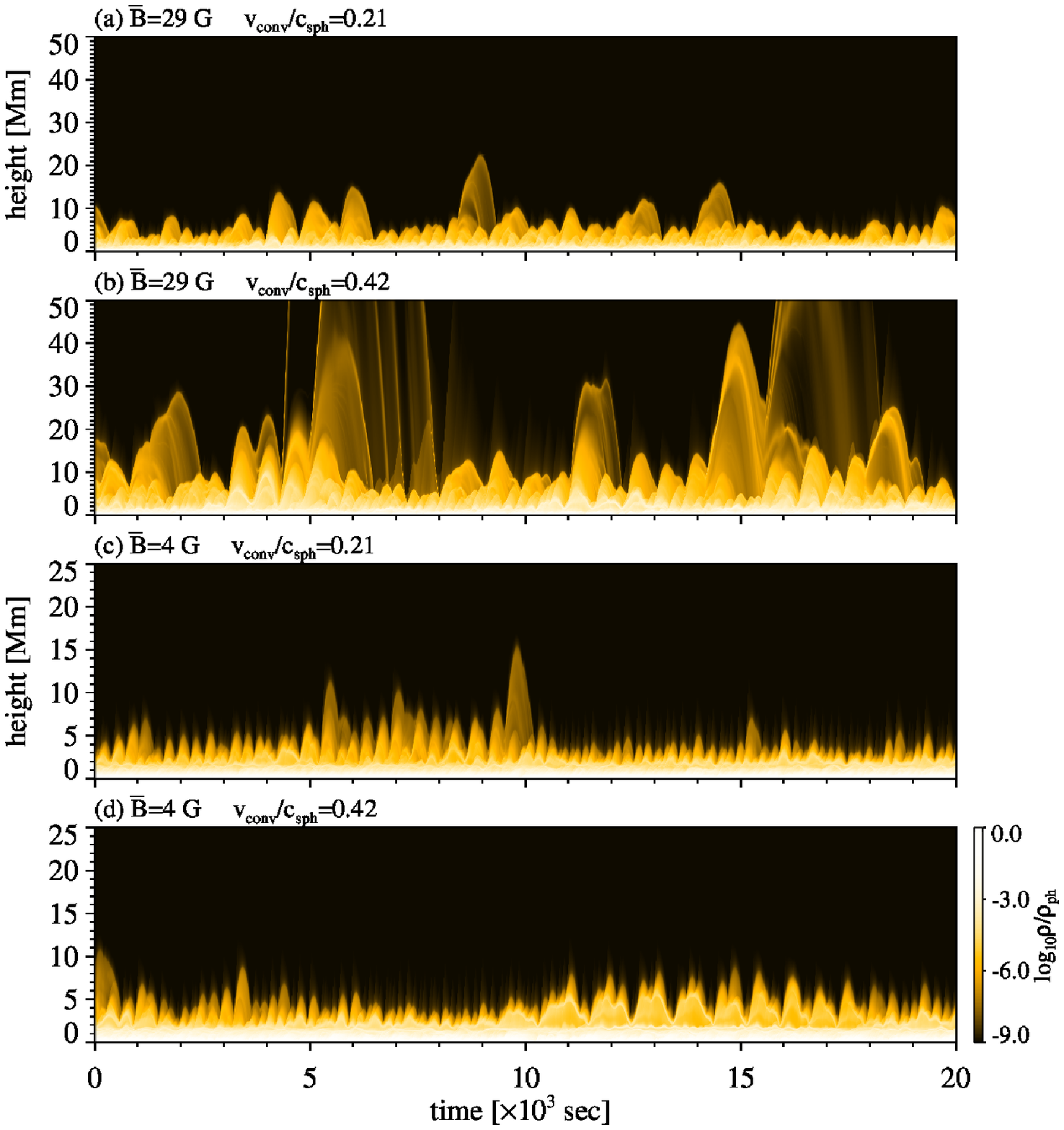}
    \caption{The timeslice diagrams of the mass density in the lower atmosphere. Note that the scale of height used in (a) and (b) is twice as large as that in (c) and (d). The top of chromosphere ($\rho/\rho_{\rm ph}\sim10^{-7}$) shows the upward and downward motion which corresponds to the spicule dynamics. The dependence of spicule dynamics on $\overline{B}$ and $v_{\rm conv}$ is clearly seen in these panels. The panels (a) and (b) show the results in the cases of $\overline{B}=29$ G and $v_{\rm conv}/c_{s\rm ph}=0.21,0.42$, respectively. The panels (c) and (d) correspond to the cases of $\overline{B}=4$ G.
    }
    \label{fig:fig05.eps}
  \end{center}
\end{figure*}
\begin{figure}
  \begin{center}
    \epsscale{1.2}  
    \plotone{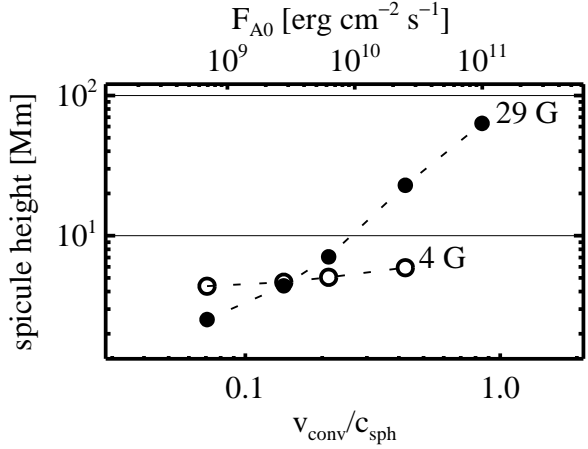}
    \caption{
      The average spicule height as a function of the velocity amplitude on the photosphere. The filled (open) circles correspond to the simulation results in the case of $\overline{B}=29$ G (4 G).
    }
    \label{fig:fig06.eps}
  \end{center}
\end{figure}
\begin{figure}
  \begin{center}
    \epsscale{1.2}  
    \plotone{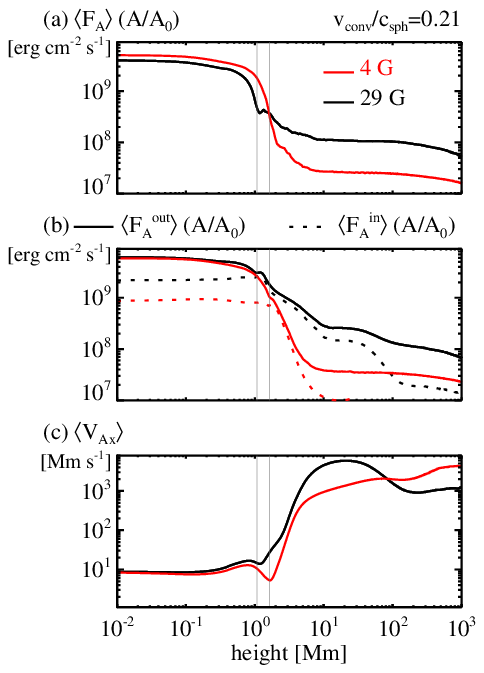}
    \caption{The dependence of transmissivity of the Alfv\'en waves on different $\overline{B}$. $v_{\rm conv}/c_{s\rm ph}=0.21$. The black and red lines show the results in the cases of $\overline{B}=29$ and 4 G. Panel (a): Poynting flux by magnetic tension force ($-B_\phi v_\phi B_x/(4\pi)$) normalized by the cross-section of magnetic flux tube. Panel (b): outward (solid lines) and inward (dashed lines) Poynting flux by magnetic tension. Panel (c): temporally averaged profile of Alfv\'en speed. The vertical gray lines correspond to the merging height $H_m=8,12H_{\rm ph}$.}
    \label{fig:fig07.eps}
  \end{center}
\end{figure}
Figure \ref{fig:fig05.eps} shows the timeslice diagrams of the mass density in the lower atmosphere. The top of chromosphere ($\rho/\rho_{\rm ph}\sim10^{-7}$) shows the upward and downward motion representing the spicule dynamics. Figure \ref{fig:fig05.eps} (a) and (b) are the results in the cases of $\overline{B}=29$ G for $v_{\rm conv}/c_{s\rm ph}=0.21$ and 0.42, respectively.

The height of the spicule becomes taller with a larger $v_{\rm conv}/c_{s\rm ph}$. On the other hand, the height of the spicule in the $\overline{B}=4$ G case is less dependent on $v_{\rm conv}$, as shown in Figure \ref{fig:fig05.eps} (c) and (d). The average spicule height, as a function of $v_{\rm conv}$, is summarized in Figure \ref{fig:fig06.eps}. The line styles and symbols are the same as those used in Figure \ref{fig:fig04.eps}. The spicule height is measured by tracking the isothermal contour of $4\times10^4$ K, the typical temperature of the transition layer \citep{2011ApJ...743..142H,2015ApJ...812L..30I}. By fitting the oscillatory pattern of the isothermal contour with the trajectories of the Lagrange particles, the individual spicules are identified, which enables us to do statistical analysis.

A common feature can be confirmed in the behaviors of the wind's mass loss rate (Figure \ref{fig:fig04.eps}) and the average spicule height (Figure \ref{fig:fig06.eps}). The spicule becomes monotonically taller with a larger $v_{\rm conv}$ in the $\overline{B}=29$ G case, while in $\overline{B}=4$ G case, it is almost independent of $v_{\rm conv}$.

The less dependence of the simulated solar wind on $v_{\rm conv}$ implies a significant wave damping below the transition layer, i.e., in the chromosphere. The difference in the spicule dynamics between $\overline{B}=4$ and 29 G also suggests the propagation of a chromospheric shock wave is qualitatively affected by the parameter $\overline{B}$. These possibilities are further investigated in the following section.

\section{ANALYSES}
\subsection{Poynting Flux by Alfv\'en Waves}
\label{sec:Poynting_flux_by_alfven_wave}
\begin{figure}
  \begin{center}
    \epsscale{1.2}  
    \plotone{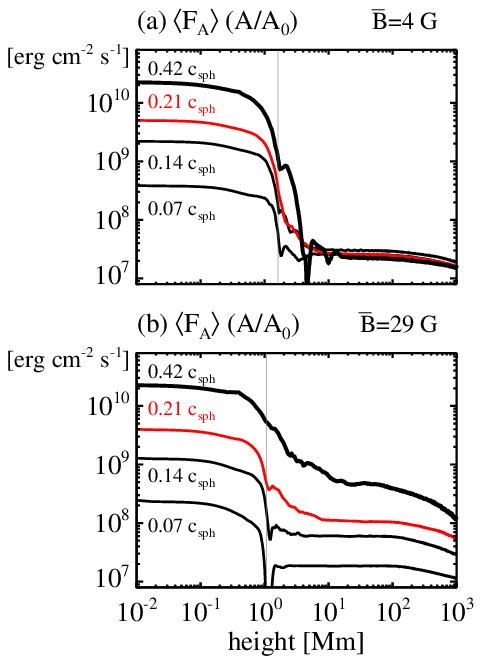}
    \caption{The dependence of transmissivity of Alfv\'en waves on $v_{\rm conv}$ in the case of $\overline{B}=4$ G (panel (a)) and 29 G (panel (b)). Each profile represents the Poynting flux of the magnetic tension force normalized by the cross section of the magnetic flux tube. The thickest black line shows the simulation result with $v_{\rm conv}/c_{s\rm ph}=0.42$ while the thick red and black lines show the results with $v_{\rm conv}/c_{s\rm ph}=0.21,0.14$. The thin line corresponds to $v_{\rm conv}/c_{s\rm ph}=0.07$. Here, $c_{s\rm ph}=\sqrt{\gamma R_gT_{\rm eff}/\mu_{\rm ph}}$ is the adiabatic sound speed on the photosphere.}
    \label{fig:fig08.eps}
  \end{center}
\end{figure}

To investigate the energy transfer by Alfv\'en waves, the time-averaged Poynting flux of the magnetic tension force ($F_A=-B_\phi v_\phi B_x/(4\pi)$) is plotted as a function of height in Figure \ref{fig:fig07.eps}. The black and red lines correspond to the results in the cases of $\overline{B}=29$ and 4 G, respectively. Although the velocity amplitude on the photosphere is fixed at $v_{\rm conv}/c_{s\rm ph}=0.21$, $F_A$ below 1 Mm in the $\overline{B}=4$ G case is slightly larger than that in the $\overline{B}=29$ G case. This is caused by the reflection of the Alfv\'en waves at the merging height. The energy flux of the reflected (inward) Alfv\'en waves is plotted in Figure \ref{fig:fig07.eps} (b) using dotted lines, where $F_A^{\rm out,in}={1\over4}\rho z_{\rm out,in}^2V_{Ax}$, $F_A=F_A^{\rm out}-F_A^{\rm in}$. As seen in this plot, the inward Alfv\'en waves below 1 Mm comes mainly from the merging height, above which the Alfv\'en speed exponentially increases (Figure \ref{fig:fig07.eps} (c)). The energy flux of the inward Alfv\'en waves below 1 Mm is, therefore, related to the outward energy flux at the merging height. This leads to the smaller net energy flux when the merging height is lower.

The most remarkable feature in Figure \ref{fig:fig07.eps} (a) is the significant decrease in the energy flux around the transition layer in the case of $\overline{B}=4$ G (red line). Figure \ref{fig:fig08.eps} (a) shows the dependence of the $F_A$ height-profile on $v_{\rm conv}$ in the case of $\overline{B}=4$ G. Although a larger $v_{\rm conv}$ produces larger $F_A$ on the bottom boundary ($F_{A0}=6\times10^8-2\times10^{10}$ erg cm$^{-2}$ s$^{-1}$ for $v_{\rm conv}/c_{s\rm ph}=0.07-0.42$), the transmitted energy fluxes into the corona do not show the significant increase from $F_A\sim 10^5$ erg cm$^{-2}$ s$^{-1}$ ($F_A(A/A_0)\sim$ a few $\times10^7$ erg cm$^{-2}$ s$^{-1}$). In other words, the additional energy input associated with larger $v_{\rm conv}$ is completely lost below the transition layer. This cannot be seen in the case of $\overline{B}=29$ G. Figure \ref{fig:fig08.eps} (b) shows that a larger energy input from the photosphere always leads to larger transmitted energy flux when $\overline{B}=29$ G.


\begin{figure*}
  \begin{center}
    \epsscale{1}
    \plotone{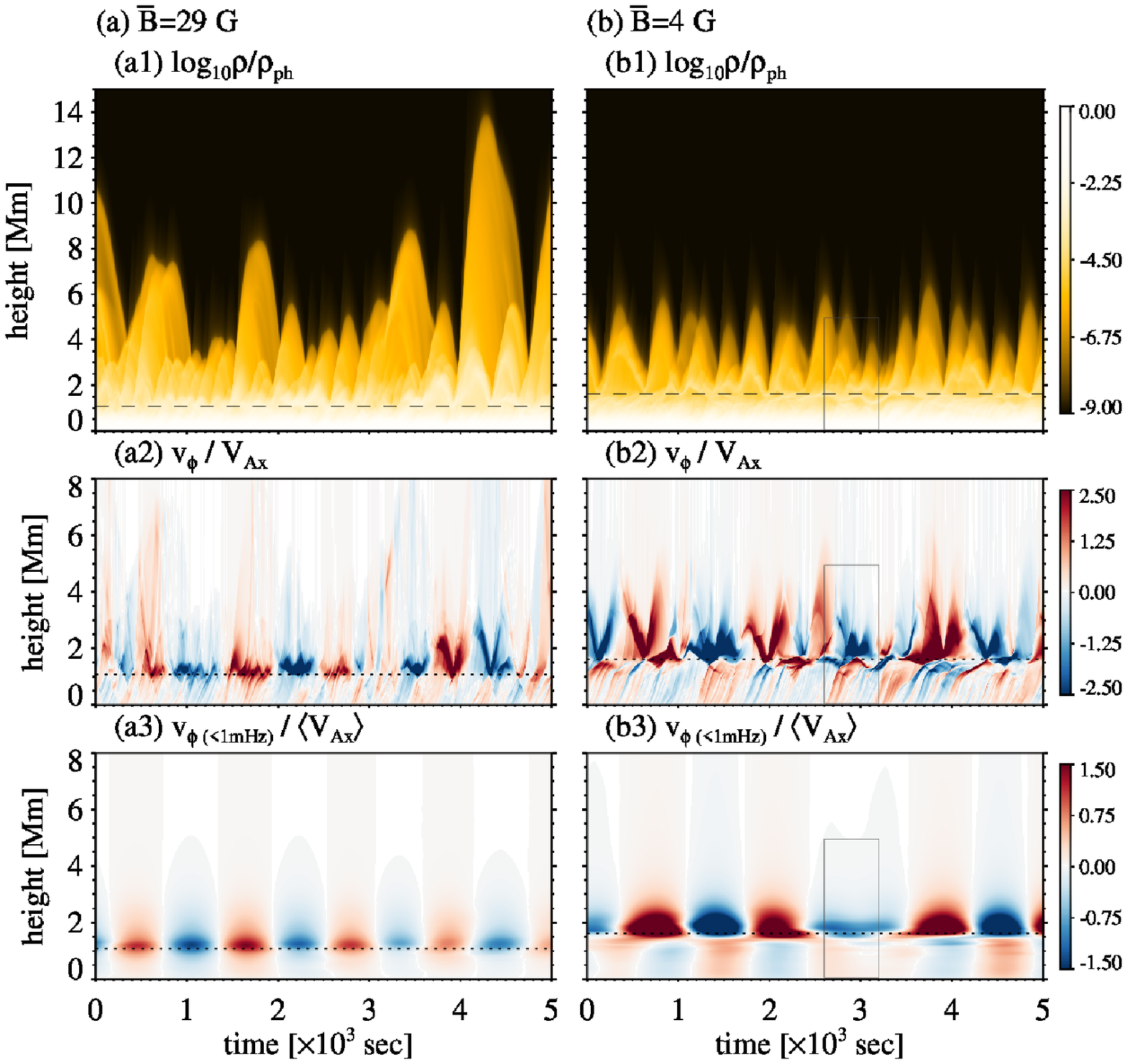}
    \caption{The twisting motion of the magnetic flux tube and its dependence on the merging height. The left and right columns show the results in the cases of $\overline{B}=29$ and 4 G, respectively. $v_{\rm conv}/c_{s\rm ph}=0.21$. The merging heights are indicated with the horizontal dashed lines. Panels (a1) and (b1): the timeslice diagram of density. Panels (a2) and (b2): the nonlinearity of toroidal velocity $v_\phi$ with respect to Alfv\'en speed $B_x/\sqrt{4\pi\rho}$. Panels (a3) and (b3): the nonlinearity of the low frequency component of toroidal velocity with respect to Alfv\'en speed. The gray rectangle area corresponds to the frame of Figure \ref{fig:fig12.eps}.}
    \label{fig:fig09.eps}
  \end{center}
\end{figure*}

\subsection{Alfv\'en Waves in the Chromosphere}
\begin{figure*}
  \begin{center}
    \epsscale{1}
    \plotone{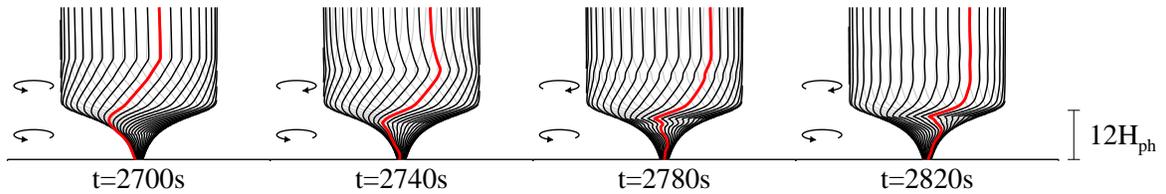}
    \caption{The schematic drawing of flux tube motion in the case of $\overline{B}=4$ G. Note that $t=0$ s corresponds to the same as used in Figure \ref{fig:fig09.eps} (b1) (b2) (b3). This time range is within the gray rectangle in Figure \ref{fig:fig09.eps}.}
    \label{fig:fig10.eps}
  \end{center}
\end{figure*}
\begin{figure*}
  \begin{center}
    \epsscale{1}
    \plotone{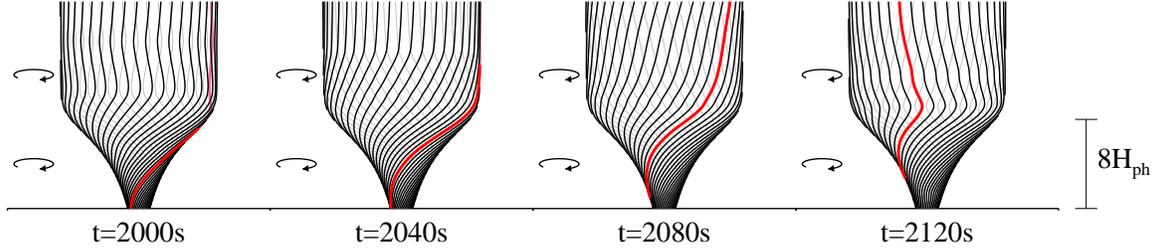}
    \caption{The schematic drawing of flux tube motion in the case of $\overline{B}=29$ G. Note that $t=0$ s corresponds to the same as used in Figure \ref{fig:fig09.eps} (a1) (a2) (a3).}
    \label{fig:fig11.eps}
  \end{center}
\end{figure*}
\begin{figure*}
  \begin{center}
    \epsscale{1.2}
    \plotone{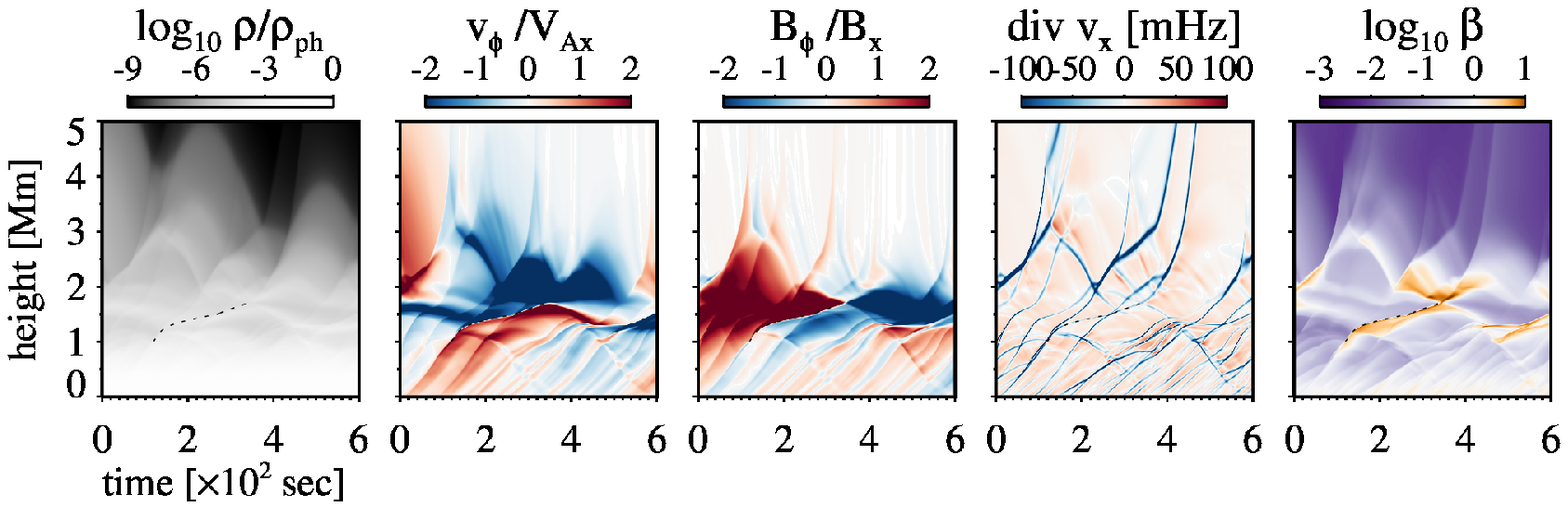}
    \caption{The timeslice diagram of $B_\phi/B_x$, div $v_x$, and plasma $\beta$ in the chromosphere, showing the highly sheared toroidal magnetic field with strong compression. The dashed line represents the propagation of the intermediate shock. $t=0$ s in these diagrams corresponds to $t=2600$ s in Figure \ref{fig:fig09.eps} (The time range of this diagram corresponds to the gray rectangle in Figure \ref{fig:fig09.eps}.).}
    \label{fig:fig12.eps}
  \end{center}
\end{figure*}
In the previous subsection, it was determined that the transmission of energy flux into the corona is limited to $\sim10^5$ erg cm$^{-2}$ s$^{-1}$ when the merging height is higher ($\overline{B}=4$ G). This suggests that the Alfv\'en waves cannot be responsible for a larger Poynting flux across the chromosphere than a certain upper limit in the case of $\overline{B}=4$ G. Therefore, how the oscillations of toroidal velocity and magnetic field depend on the poloidal magnetic field configuration in the chromosphere were investigated.

Figure \ref{fig:fig09.eps} shows the twisting motion of the magnetic flux tube in the chromosphere. Figures \ref{fig:fig09.eps} (a1) and (b1) show the timeslice diagram of density in the lower atmosphere when $\overline{B}=29$ and 4 G, respectively. Figures \ref{fig:fig09.eps} (a2) and (b2) show the nonlinearity of Alfv\'en wave amplitude. Because of the weaker $\overline{B}$, $v_\phi/V_{Ax}$ are higher in the $\overline{B}=4$ G case. In addition, the toroidal velocity above and below the merging height (horizontal dashed lines) often have the opposite sign when $\overline{B}=4$ G. Such an anti-phase oscillation is rarely seen when $\overline{B}=29$ G. This difference is more clearly seen in Figures \ref{fig:fig09.eps} (a3) and (b3). These panels show the comparison of the low frequency component of the $v_\phi$ oscillation ($\nu<$1mHz). The anti-phase oscillation mentioned above appears in Figure \ref{fig:fig09.eps} (b3).

Figures \ref{fig:fig10.eps} and \ref{fig:fig11.eps} depict the typical time sequence of the magnetic field lines in the cases of $\overline{B}=4$ and 29 G, respectively. When the merging height is low and $\overline{B}$ is large, the upper part of the flux tube above the merging height is twisted as its lower part rotates (Figure \ref{fig:fig11.eps}). On the other hand, Figure \ref{fig:fig10.eps} shows that the upper part of the flux tube is counter-rotating against lower part, thereby causing the formation of the break of the magnetic field line. The close-up view around such a break of magnetic field line is shown in Figure \ref{fig:fig12.eps}, which corresponds to the rectangle area in Figure \ref{fig:fig09.eps}. The break of the magnetic field line is represented by the dashed line in this figure, which agrees with the characteristics at $v_x+B_x/\sqrt{4\pi\rho}$. Figure \ref{fig:fig09.eps} shows that this signature appears transiently and is associated with a compression that is strong enough to significantly enhance the plasma $\beta$ in the downstream. The break of the magnetic field line is, therefore, identified as the intermediate shock.

\subsection{Slow / Fast Shocks in the Chromosphere}
\label{sec:Slow_Fast_Shocks_in_the_Chromosphere}
The previous subsections revealed that the energy transfer by the Alfv\'en waves are restricted in the case of the weak magnetic field ($\overline{B}=4$ G). Aside from such a nearly incompressible wave, the propagation of magnetoacoustic shocks, including slow and fast shocks, are possibly dependent on the magnetic field configuration in the chromosphere. In fact, Figure \ref{fig:fig06.eps} shows the dependence of average spicule height on $v_{\rm conv}$ changes in accordance with the magnetic field strength $\overline{B}$. For a comprehensive discussion, we investigated the propagation of slow and fast shocks.

The relatively strong compressible wave can be distinguished as the propagating spiky signatures with $-\partial_xv_x>0$. After tracing these signatures, the Alfv\'en Mach number of the shock wave ($M_A$) is calculated using the following formula (the derivation is described in Appendix \ref{sec:Mach_number_of_MHD_shock_wave}):
\begin{equation}
  M_A=-{1\over V_A}\left({\partial v_x\over\partial x}\right)^{-1}\left\{{1\over\rho}{\partial p_{\rm tot}\over\partial x}-{\partial\over\partial x}\left({GM_\odot\over r}\right)\right\}
  \label{eq:mach_number}
\end{equation}
where $p_{\rm tot}=p+B_\phi^2/(8\pi)$. By expressing the fast and slow mode Mach numbers with $M_f=M_AV_A/V_{\rm fast}$ and $M_s=M_AV_A/V_{\rm slow}$ where $V_{\rm fast}$ and $V_{\rm slow}$ are the fast and slow mode speeds, the detected shock is specified as the fast shock when $|M_f-1|<|M_s-1|$, or, otherwise, the slow shock. This classification is justified when both fast and slow shocks are relatively weak, i.e., $M_f\sim1$ and $M_s\sim1$. By counting the fast (slow) shocks with $M_f$ ($M_s$) propagating around the mass density $\rho$ in the stratified atmosphere, the distribution function of $M_f$ or $M_s$, with respect to $\rho$, is defined as follows:
\begin{equation}
  {dN\over d\log_{10}\rho dM}(\rho,M)={dN(\rho\mbox{ \footnotesize$\in[\rho_i,\rho_{i+1}]$},M\mbox{ \footnotesize$\in[M_j,M_{j+1}]$})\over(\log_{10}\rho_{i+1}-\log_{10}\rho_i)(M_{j+1}-M_j)}
  \label{eq:dist_func_shock}
\end{equation}
where $dN(\rho,M)$ is the expected number of shocks characterized with ($\rho,M$) in one snapshot and the subscriptions $i$ and $j$ represent the discretization.

\begin{figure*}
  \begin{center}
    \epsscale{1.2}
    \plotone{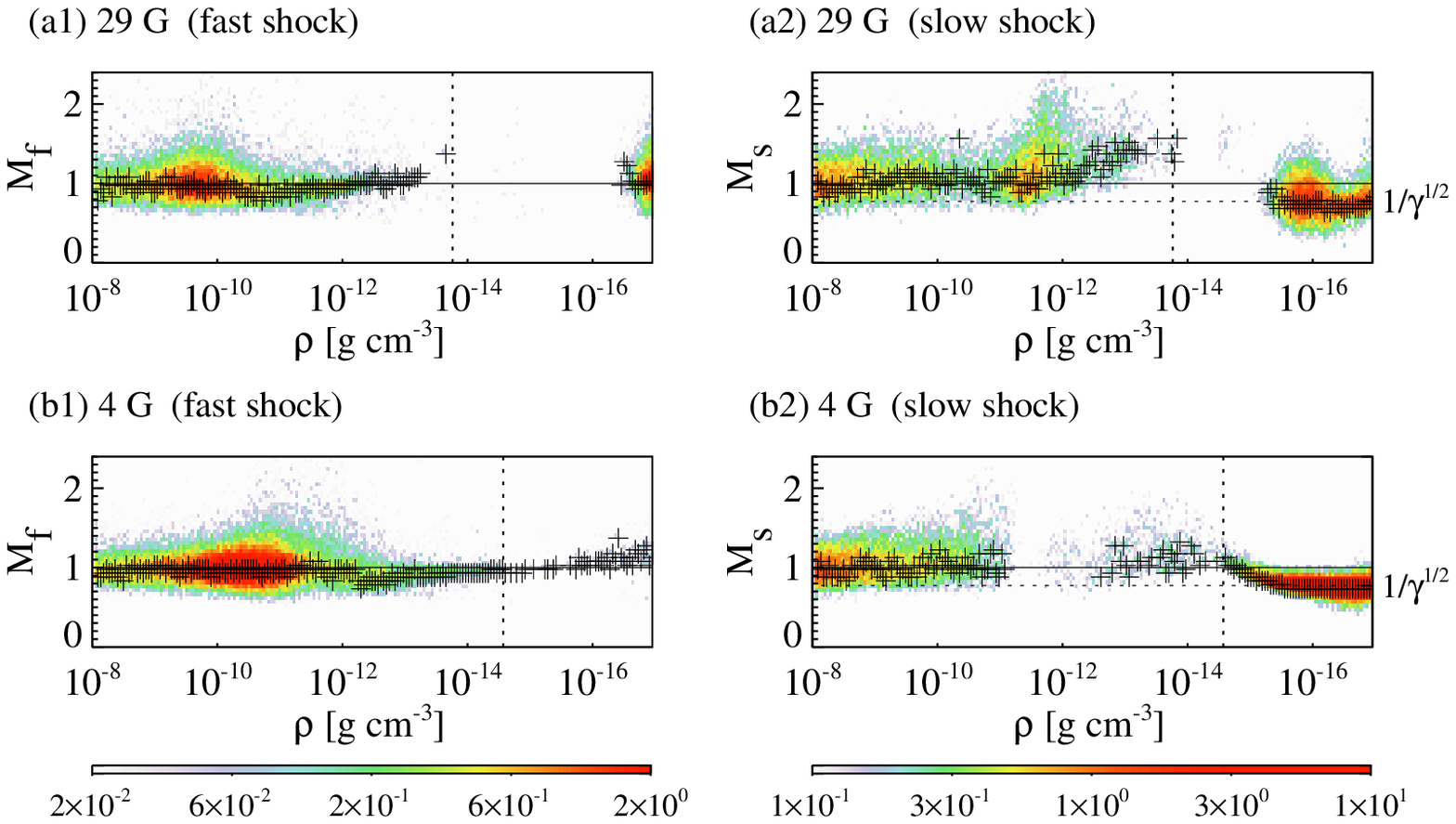}
    \caption{The distribution functions of the Mach number of the fast or slow shocks with respect to the atmospheric mass density $\rho$. The color shows the quantities defined by Eq. \ref{eq:dist_func_shock}; the expected number of shocks found in one snapshot. The cross symbol represents the most frequently appearing Mach number in each bin of $\rho$. The panels (a1) and (a2) show the analysis result for the case of $\overline{B}=29$ G while (b1) and (b2) correspond to the $\overline{B}=4$ G case. $v_{\rm conv}/c_{s\rm ph}=0.21$. The vertical dotted line in each panel corresponds to the mean mass density at the transition layer. The horizontal dotted lines in the panels (a2) and (b2) correspond to $M_s=1/\sqrt{\gamma}$.
    }
    \label{fig:fig13.eps}
  \end{center}
\end{figure*}

Figure \ref{fig:fig13.eps} shows the distribution functions calculated from the simulation results in the cases of $\overline{B}=29$ G (upper panels) and 4 G (lower panels). $v_{\rm conv}/c_{s\rm ph}$ is fixed at 0.21. The vertical dotted line in each panel corresponds to the mean mass density at the transition layer. The distribution around the transition layer is artificially sparse in all panels. This is because the shock crossing the transition layer is hardly detected in this analysis (we used the time series data over 50000 sec with an interval of 4 sec. This interval is much longer than that of the shock crossing timescale across the transition layer). The cross symbol represents the most frequently appearing Mach number in each bin of $\rho$. Thus, the gradual rise of cross symbols seen around $\rho\gtrsim10^{-12}$ g cm$^{-3}$ in Figure \ref{fig:fig13.eps} (a2) shows the growth of the chromospheric slow shock. Note that the coronal slow shocks in Figures \ref{fig:fig13.eps} (a2) and (b2) concentrate on $M_s\sim1/\sqrt{\gamma}$ rather than $M_s\sim1$. This is bacause we calculated $M_s$ by assuming $\gamma=5/3$ without any considerations of non-adiabatic effects. The phase speed of slow shock in the corona tend to be the isothermal sound speed $\sim\sqrt{p/\rho}$ due to the strong heat conduction. This leads to the underestimation of $M_s$ by a factor of $\sim1/\sqrt{\gamma}$ in the corona.

The most remarkable feature in this figure is that the slow shock vanishes around $\rho\sim10^{-11}$ g cm$^{-3}$ in Figure \ref{fig:fig13.eps} (b2). $\rho=10^{-11}$ g cm$^{-3}$ is three orders of magnitude higher than the mass density at the transition layer and roughly corresponds to the mass density around the merging height. Therefore, this disappearance of slow shock is not related to the abovementioned artificial sparse distribution around the transition layer. Instead, it is implied that the slow shock can be evanescent in the chromosphere when the magnetic field is weak.

\section{DISCUSSION}
\subsection{Intermediate Shock in the Chromosphere}
\label{sec:Intermediate_Shock_in_the_Chromosphere}
\begin{figure}
  \begin{center}
    \epsscale{1.2}  
    \plotone{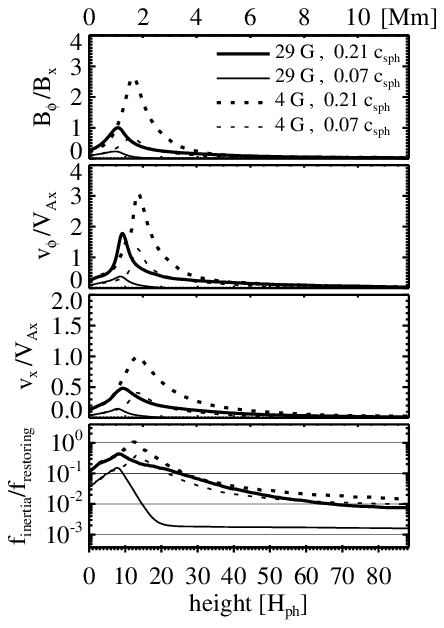}
    \caption{
      The nonlinearity of $B_\phi$, $v_\phi$, $v_x$ in lower atmosphere. The solid and dashed lines show the results in the cases of $\overline{B}=29$ and 4 G, respectively. The thick and thin lines show the results in the cases of $v_{\rm conv}/c_{s\rm ph}=0.21$ and 0.07, respectively.
    }
    \label{fig:fig14.eps}
  \end{center}
\end{figure}
\begin{figure*}
  \begin{center}
    \epsscale{1}
    \plotone{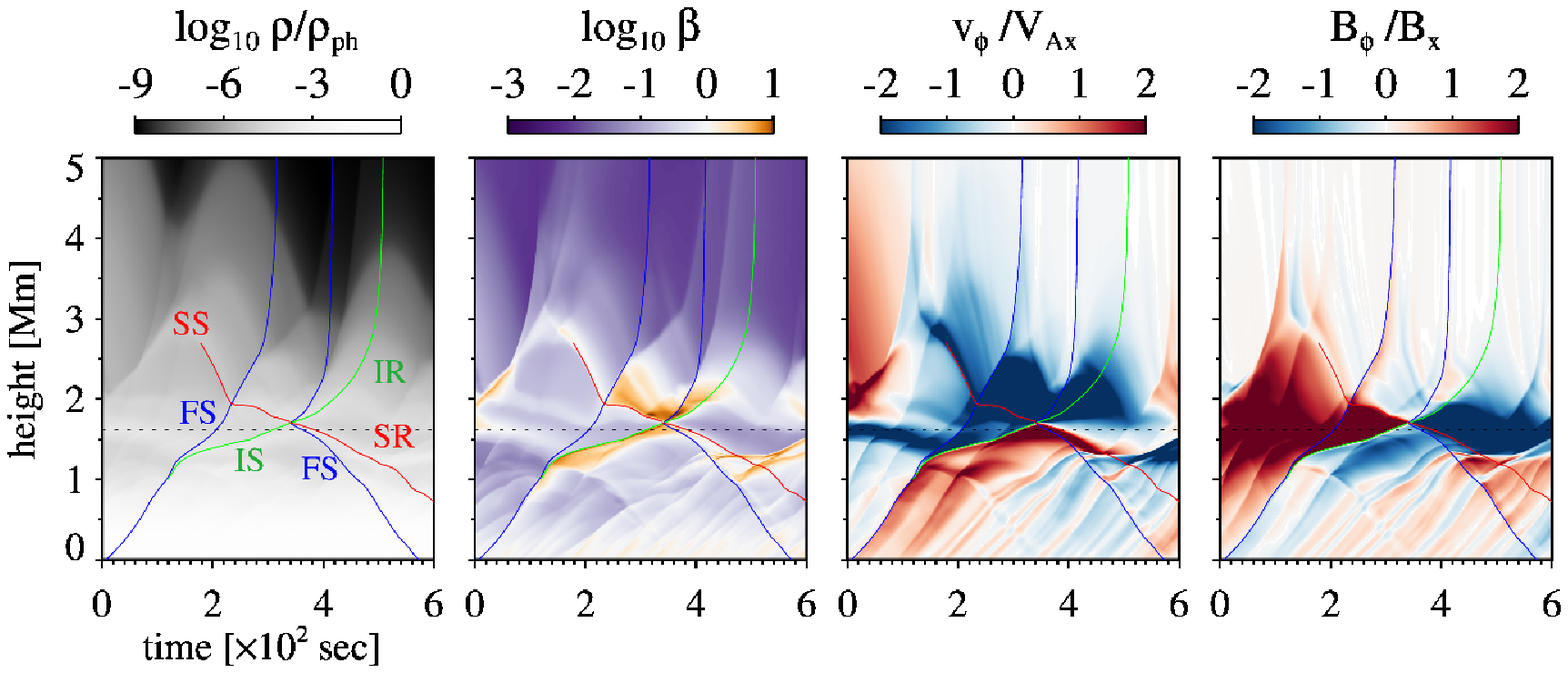}
    \caption{The typical scene of formation of the intermediate shock which deviates from the fast shock. $t=0$ s in these diagrams corresponds to $t=2600$ s in Figure \ref{fig:fig09.eps}.
      The intermediate shock immediately interacts with the downward slow shock and results in the intermediate rarefaction wave with the negative Pointing flux. ``SS'', ``FS'', ``IS'', ``IR'', and ``SR'' in the leftmost panel stand for the slow shock, fast shock, intermediate shock, and intermediate rarefaction wave, slow rarefaction wave, respectively. The colored lines correspond to the trajectories of characteristics. The horizontal dashed line represents the merging height $H_m=12H_{\rm ph}$.}
    \label{fig:fig15.eps}
  \end{center}
\end{figure*}

We discuss here the causal relationship between the high nonlinearity of the Alfv\'en waves in the chromosphere and the limit on the energy transmission into the corona (section \ref{sec:Poynting_flux_by_alfven_wave}).

Figure \ref{fig:fig14.eps} shows the temporally averaged profiles of nonlinearity regarding $B_\phi$, $v_\phi$, and $v_x$. Note that we abbreviated $\langle(B_\phi/B_x)^2\rangle^{1/2}$, $\langle(v_\phi/V_{Ax})^2\rangle^{1/2}$, $\langle(v_x/V_{Ax})^2\rangle^{1/2}$ into $B_\phi/B_x$, $v_\phi/V_{Ax}$, $v_x/V_{Ax}$. $B_\phi/B_x$ and $v_\phi/V_{Ax}$ represent the nonlinearity of Alfv\'en waves. The maximum level of nonlinearity is always found around the merging height. The higher merging height is responsible for the higher maximum nonlinearity of torsional ($B_\phi$ and $v_\phi$) and longitudinal ($v_x$) oscillation. In particular, the high level of nonlinearity of $v_x$ corresponds to the large inertia of the magnetic flux tube.

By using the mass conservation (Eq. \ref{eq:mass}) and the poloidal magnetic flux conservation (Eq. \ref{eq:mag_pol}), the toroidal component of the equation of motion (Eq. \ref{eq:mom_tor}) can be expressed as below.
\begin{equation}
  {\partial v_\phi\over\partial t}+{v_x\over\sqrt{A}}{\partial(\sqrt{A} v_\phi)\over\partial x}-{B_x\over4\pi\rho\sqrt{A}}{\partial(\sqrt{A} B_\phi)\over\partial x}=0
  \label{eq:eq_eom_tor}
\end{equation}
The variables in the above equation are the same as used in section \ref{sec:basic_equations}. The second term represents the inertia term. It can be competitive against the Lorentz force (the third term) when the longitudinal oscillation is highly nonlinear.
The last panel of Figure \ref{fig:fig14.eps} shows the ratio of the temporally averaged absolute value of inertia term in relation to that of the restoring term in Eq. \ref{eq:eq_eom_tor}. Here, we define the following:
\begin{equation}
  f_{\rm inertia}=\left\langle\left|{v_x\over\sqrt{A}}{\partial(\sqrt{A}v_\phi)\over\partial x}\right|\right\rangle
  \label{eq:eq_f_inetia}
\end{equation}
\begin{equation}
  f_{\rm restoring}=\left\langle\left|{B_x\over4\pi\rho\sqrt{A}}{\partial(\sqrt{A}B_\phi)\over\partial x}\right|\right\rangle
  \label{eq:eq_f_restoring}
\end{equation}
The low ratio in the corona means that the Alfv\'en waves can propagate without a significant nonlinear effect while the ratio around the unity implies that the wave propagation is strongly affected by the inertia term. In the case of the higher merging height ($\overline{B}=4$ G), the ratio reaches the unity around the merging height, as shown with the red thick line in Figure \ref{fig:fig15.eps}. Due to this large inertia of the magnetic flux tube, the rotation of the upper part of the flux tube cannot be restored so easily by the twisting motion injected from the photosphere. That would result in the anti-phase oscillation between the upper and lower parts of the flux tube (Figure \ref{fig:fig09.eps}). The highly sheared torsional flow and nonlinear longitudinal oscillation can cause the ``fracture'' of the flux tube, i.e., the formation of the intermediate shock (Figure \ref{fig:fig10.eps}). Once the intermediate shock is formed in the chromosphere, the Poynting flux associated with it hardly transmits into the corona. That is because the intermediate shock easily interacts with the slow and fast shocks or contact discontinuity, including the transition layer itself. Among these interactions, the collision of intermediate shock with the transition layer results in the transmitted waves composed of fast rarefaction wave and slow shock. Since both of them have negative Poynting fluxes, the magnetic energy transferred by the chromospheric intermediate shock is, in this sense, confined below the transition layer until its dissipation.

\begin{figure*}
  \begin{center}
    \epsscale{1}
    \plotone{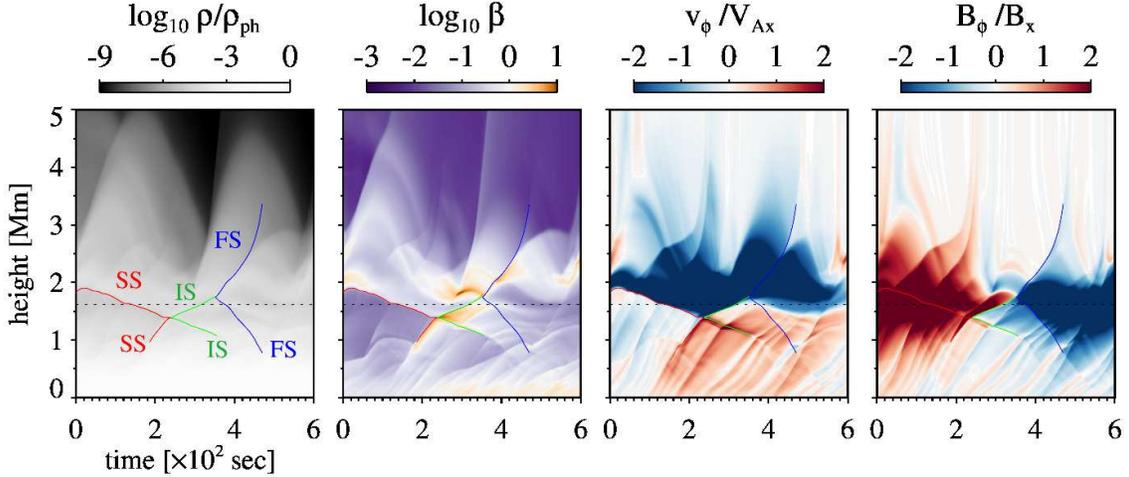}
    \caption{The typical scene of formation of the intermediate shock which results from the head-on collision of slow shocks. $t=0$ s in these diagrams corresponds to $t=1100$ s in Figure \ref{fig:fig09.eps}. 
      ``SS'', ``FS'', and ``IS'' in the leftmost panel stand for the slow shock, fast shock, and intermediate shock, respectively. The colored lines correspond to the trajectories of characteristics. The horizontal dashed line represents the merging height $H_m=12H_{\rm ph}$.}
    \label{fig:fig16.eps}
  \end{center}
\end{figure*}

\begin{figure*}
  \begin{center}
    \epsscale{1}
    \plotone{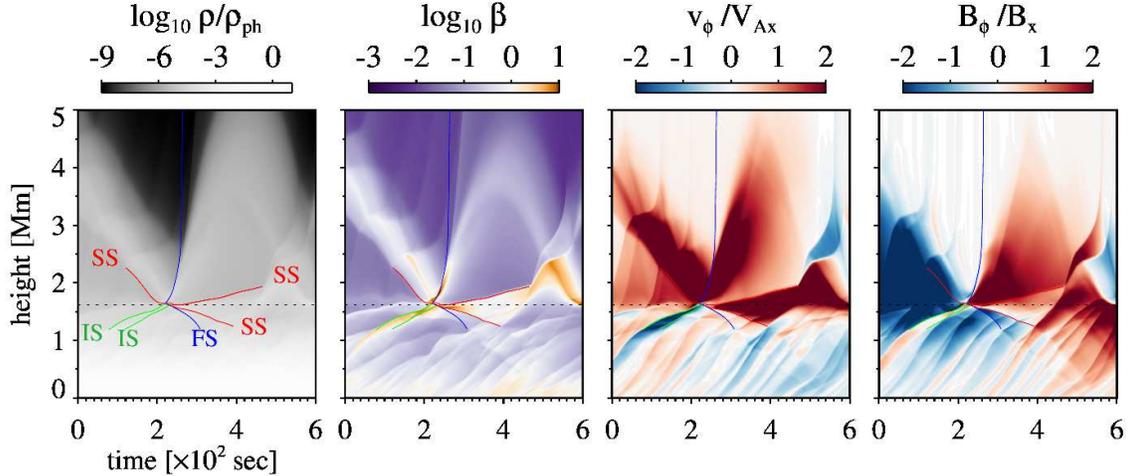}
    \caption{The typical scene of rapid dissipation of the intermediate shock. $t=0$ s in these diagrams corresponds to $t=400$ s in Figure \ref{fig:fig09.eps}. The collision between the sequences of upward intermediate shocks with the downward slow shock leads to the bidirectional slow shocks. ``SS'', ``FS'', and ``IS'' in the leftmost panel stand for the slow shock, fast shock, and intermediate shock, respectively. The colored lines correspond to the trajectories of characteristics. The horizontal dashed line represents the merging height $H_m=12H_{\rm ph}$.}
    \label{fig:fig17.eps}
  \end{center}
\end{figure*}

Figures \ref{fig:fig15.eps} and \ref{fig:fig16.eps} show the examples on the formation of chromospheric intermediate shock. In Figure \ref{fig:fig15.eps}, the intermediate shock deviates from the fast shock at around $t=120$ s and collides with the downward slow shock at around $t=340$ s. Although the upward fast shock generated by this collision has a positive Poynting flux, the other resultant waves, including the upward intermediate rarefaction wave, transport the magnetic energy downward. The formation of the intermediate shock in Figure \ref{fig:fig16.eps} is a result from the head-on collision of upward and downward slow shocks, which is associated with the encounter of large shear flow. The upward intermediate shock finally becomes the bidirectional fast shocks after the interaction with the other waves. The dissipation of the intermediate shock is clearly exemplified in Figure \ref{fig:fig17.eps}. In this scene, the interaction between the sequence of intermediate shocks and the downward slow shock results in the bidirectional slow shocks. As a result, the highly sheared magnetic field line is rapidly relaxed and the super-Alfv\'enic torsional flow is generated.

\subsection{Wave Nonlinearity in the Chromosphere}
\label{sec:Wave_Nonlinearity_in_the_Chromosphere}
\begin{figure*}
  \begin{center}
    \epsscale{1}
    \plotone{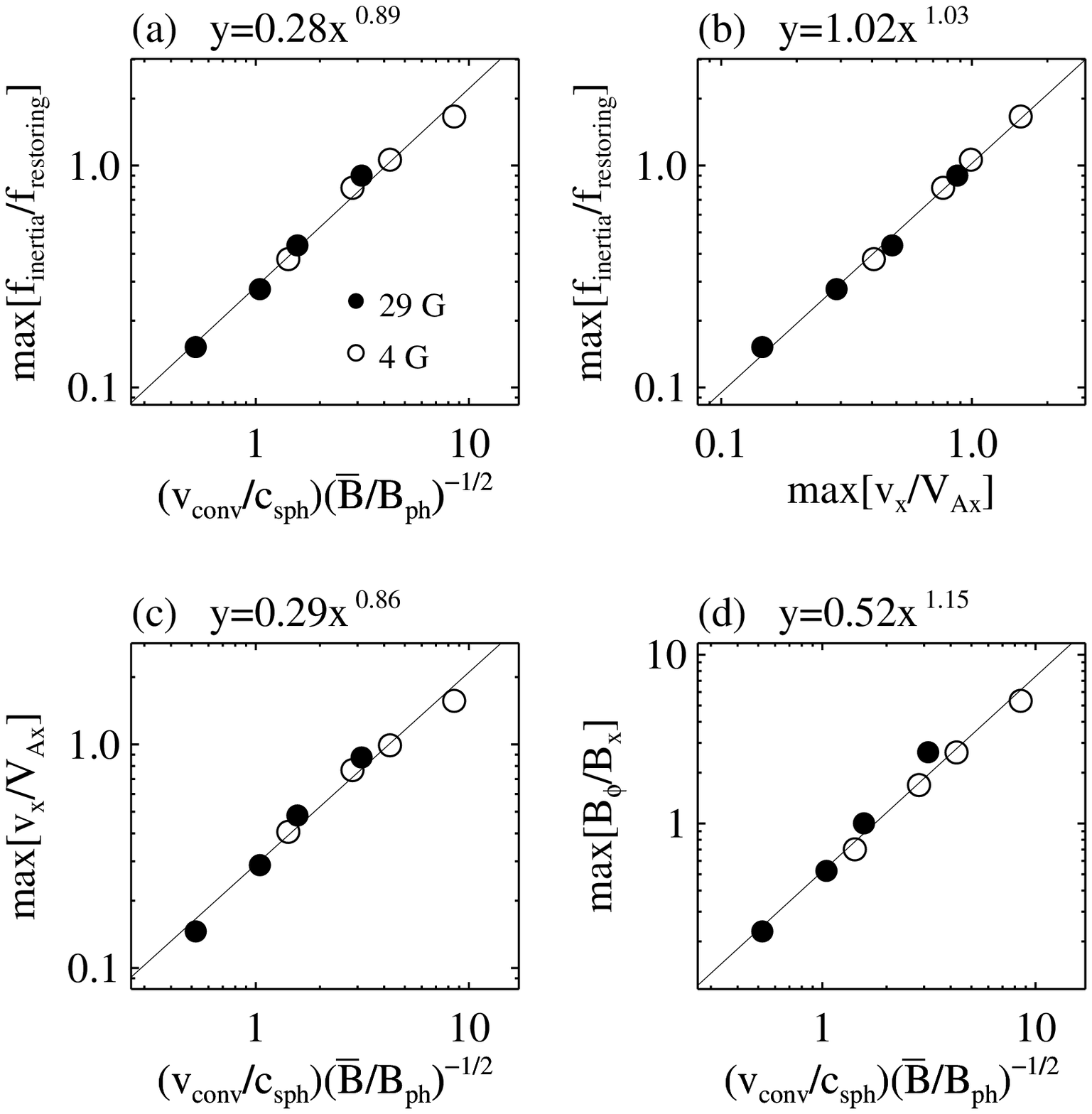}
    \caption{The scaling relation between $v_{\rm conv}/c_{s\rm ph}$ and wave nonlinearity in the chromosphere. The plotted quantities correspond to the maximum values of the profiles shown in Figure \ref{fig:fig14.eps}. The styles of squares or circles are the same as those used in Figure \ref{fig:fig04.eps}.}
    \label{fig:fig18.eps}
  \end{center}
\end{figure*}

Figure \ref{fig:fig14.eps} shows that the wave nonlinearity such as $B_\phi/B_x$, $v_\phi/V_{Ax}$, $v_x/V_{Ax}$ is the highest around the merging height. It demonstrates that higher merging height (or weaker $\overline{B}$) and larger $v_{\rm conv}/c_{s\rm ph}$ are always associated with higher wave nonlinearity in the chromosphere. By focusing on the maximum values of the profiles plotted in Figure \ref{fig:fig14.eps}, the scaling relations between $v_{\rm conv}/c_{s\rm ph}$ and wave nonlinearity were summarized in Figure \ref{fig:fig18.eps}. 

Figure \ref{fig:fig18.eps} (a) shows that the ratio of the inertia force to the restoring force is clearly correlated to $(v_{\rm conv}/c_{s\rm ph})(\overline{B}/B_{\rm ph})^{-1/2}$, i.e.,
\begin{equation}
  f_{\rm inertia}/f_{\rm restoring}=0.28[(v_{\rm conv}/c_{s\rm ph})(\overline{B}/B_{\rm ph})^{-1/2}]^{-0.89}
  \label{eq:eq_ratio_force}
\end{equation}
This scaling is composed of the relation between $f_{\rm inertia}/f_{\rm restoring}$ with $v_x/V_{Ax}$ (Figure \ref{fig:fig18.eps} (b)) and that between $v_x/V_{Ax}$ with $(v_{\rm conv}/c_{s\rm ph})(\overline{B}/B_{\rm ph})^{-1/2}$ (Figure \ref{fig:fig18.eps} (c)). In fact, Eq. \ref{eq:eq_f_inetia} and \ref{eq:eq_f_restoring} indicate that $f_{\rm inertia}\sim v_xv_\phi/\lambda_A$ and $f_{\rm restoring}\sim B_xB_\phi/(4\pi\rho\lambda_A)$, where $\lambda_A$ is the wavelength of the Alfv\'en waves. Therefore, $f_{\rm inertia}/f_{\rm restoring}$ tends to be $v_x/V_{Ax}$ when $v_\phi\sim B_\phi/\sqrt{4\pi\rho}$. The amplitude of $v_x$ basically follows the energy flux conservation for the longitudinal wave in the isothermal atmosphere. That means $\rho v_x^2/B_x\sim$ constant and $v_x/V_{Ax}\propto B_x^{-1/2}$. By using these scaling relations, it is inferred that there is a critical $v_{\rm conv}/c_{s\rm ph}$ or $\overline{B}/B_{\rm ph}$ across which the chromosphere is too highly nonlinear such that the Lorentz force associated with Alfv\'en wave propagation ($f_{\rm restoring}$) can no longer twist the flux tube against the large inertia force ($f_{\rm inertia}$). From Eq. \ref{eq:eq_ratio_force}, we replace this critical condition of $f_{\rm inertia}\gtrsim f_{\rm restoring}$ with $0.28[(v_{\rm conv}/c_{s\rm ph})/\sqrt{\overline{B}/B_{\rm ph}}]^{-0.89}\gtrsim 1$ or
\begin{equation}
  v_{\rm conv}/c_{s\rm ph}\gtrsim4.2\sqrt{\overline{B}/B_{\rm ph}}
\end{equation}
This implies the following: first, for a given flux tube with $\overline{B}/B_{\rm ph}$, the energy input from the photosphere larger than $F_{A,\rm cr}=\rho_{\rm ph}v_{\rm conv,cr}^2V_{A\rm ph}$ does not contribute to the coronal heating. As such, 
\begin{equation}
  F_{A,\rm cr}=2.4\times10^{12}\ \mbox{erg cm$^{-2}$ s$^{-1}$} (\overline{B}/B_{\rm ph})
  \label{eq:facr_ax}
\end{equation}
When $\overline{B}=$4 G and $B_{\rm ph}=1560$ G, we find $(v_{\rm conv}/c_{s\rm ph})_{\rm cr}\sim0.21$ and $F_{A,\rm cr}=6.3\times10^9$ erg cm$^{-2}$ s$^{-1}$. Second, for a given convection velocity of $v_{\rm conv}/c_{s\rm ph}$, the magnetic flux tube with $\overline{B}/B_{\rm ph}<(\overline{B}/B_{\rm ph})_{\rm cr}=0.057(v_{\rm conv}/c_{s\rm ph})^2$ is unable to guide the magnetic energy from the lower atmosphere to the corona. When $v_{\rm conv}/c_{s\rm ph}=0.21$ and $B_{\rm ph}=1560$ G, we find $\overline{B}_{\rm cr}=$4 G.

Finally, it is notable that the wave nonlinearity of $B_\phi/B_x$ follows $B_\phi/B_x\propto(\overline{B}^{-1/2})^{1.15}$ (Figure \ref{fig:fig18.eps} (d)). This is accounted for by the energy flux conservation for the Alfv\'en waves propagating along the magnetic flux tube that expands like $B_x^2\propto\rho$. Since $\rho v_\phi^2V_{Ax}A=$ const., we find $v_\phi\propto\rho^{-1/4}$ and $B_\phi/B_x\sim v_\phi/V_{Ax}\propto \rho^{1/4}/B_x$. Thus, when $B_x^2\propto\rho$, it is obtained that $B_\phi/B_x\propto B_x^{-1/2}$. \citet{1971JGR....76.5155H} and \citet{1985PASJ...37...31S} discussed that large-amplitude Alfv\'en waves can be responsible for the longitudinal motion and derived the relationship of $v_x/V_{Ax}\propto(B_\phi/B_x)^2$. On the other hand, Figures \ref{fig:fig18.eps} (c) and (d) show $v_x/V_{Ax}=0.29[(v_{\rm conv}/c_{s\rm ph})/\sqrt{(\overline{B}/B_{\rm ph})}]^{0.86}$ and $B_\phi/B_x=0.52[(v_{\rm conv}/c_{s\rm ph})/\sqrt{(\overline{B}/B_{\rm ph})}]^{1.15}$, leading to the following scaling law:
\begin{equation}
  v_x/V_{Ax}=0.51(B_\phi/B_x)^{0.75}
\end{equation}

\subsection{Evanescence of Slow Shock in the Chromosphere}

Figure \ref{fig:fig16.eps} exhibits the formation of the intermediate shock as well as the disappearance of the upward slow shock. This reminds us of Figure \ref{fig:fig13.eps}, which shows that the slow shock is absent in the upper chromosphere when the magnetic field is weak. In addition to this head-on collision of the counter-propagating slow shocks, the head-on and rear-end collisions between the slow and intermediate shocks can disturb the upward propagation of the slow shock. These interactions would be encouraged in the highly-nonlinear chromosphere, especially when the magnetic field is weak. This is because the crossing time scale of slow shock at a speed of $\sim c_sB_x/B$ becomes longer as nonlinearity increases.

This evanescence of the slow shock in the chromosphere could result in the following two consequences about the spicule dynamics. First, the ejection speed of spicule would become smaller and less dependent on $v_{\rm conv}$. Second, less frequent slow shocks in the upper chromosphere could reduce the chromospheric temperature, leading to a shorter density scale height in the chromosphere (Appendix \ref{sec:density_scale_height}). As a result of the smaller ejection speed and shorter density scale height, the average spicule height in the weak magnetic field tends to be lower than that in the strong magnetic field.

\subsection{Comparison with Observation and Other Theoretical Studies}
\subsubsection{Solar Wind}

\begin{figure}
  \begin{center}
    \epsscale{1.2}  
    \plotone{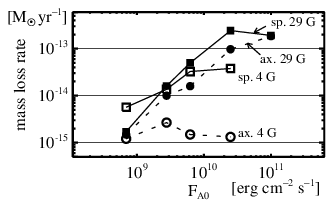}
    \caption{
      The mass loss rates of solar wind as a function of the energy input from the photosphere ($F_{A0}$). The circle (square) symbols correspond to the simulation results with the axisymmetric (local-spherically symmetric) coordinate system. The filled (open) symbols mean the results of $\overline{B}$=29 G (4 G) case.
    }
    \label{fig:fig19.eps}
  \end{center}
\end{figure}

\begin{figure}
  \begin{center}
    \epsscale{1.2}  
    \plotone{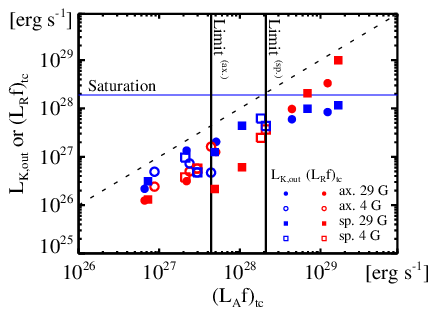}
    \caption{
      The kinetic energy flux (blue symbols) or radiative loss (red symbols) of the solar wind with respect to Poynting flux at the top of chromosphere ($T=2\times10^4$ K). This figure is analogous to Figure 8 in \citet{2013PASJ...65...98S}. The blue horizontal line corresponds to the saturation level suggested by their study (Eq. \ref{eq:saturation_level}). The relation of $y=x$ is also plotted using the dotted line. The vertical thick lines indicate the limit of transmitted Poynting flux found in our simulation with the axisymmetric (Limit$^{(\rm ax.)}$) and local-spherically symmetric coordinate systems (Limit$^{(\rm sp.)}$).
    }
    \label{fig:fig20.eps}
  \end{center}
\end{figure}
The typical fast solar wind proton flux observed around 1 AU is $\sim2\times10^8$ cm$^{-2}$ s$^{-1}$ \citep{1989ApJ...337L..49W,2010ApJ...715L.121W}, comparable to the simulated value in the stronger magnetic field case ($\sim2.1\times10^8$ cm$^{-2}$ s$^{-1}$ for $\overline{B}=29$ G), but inconsistent with that in the weaker magnetic field case ($\sim0.20\times10^8$ cm$^{-2}$ s$^{-1}$ for $\overline{B}=4$ G). As discussed in section \ref{sec:Intermediate_Shock_in_the_Chromosphere}, Poynting flux into the corona is limited to 10$^5$ erg cm$^{-2}$ s$^{-1}$ when $\overline{B}=4$ G, which causes a significantly low mass flux of solar wind. The inconsistency between the observed and simulated mass loss rates in the $\overline{B}=4$ G case is, however, easily solved by considering the polarization of the Alfv\'en waves. In the present study, we used the axisymmetric coordinate system with a linearly polarized Alfv\'en waves. The non-linear propagation of the circularly polarized Alfv\'en waves in the non-steady solar wind was simulated by \citet{2006JGRA..111.6101S} and \citet{2018ApJ...853..190S} using the local-spherically symmetric coordinate system \citep{2018ApJ...854....9S}. The differences between the axisymmetric and local-spherically symmetric coordinate systems are summarized in Appendix \ref{sec:appendix_coordinate}. We also conducted a similar parameter survey on the solar atmosphere and wind structure in the local-spherically symmetric coordinate system with the circularly polarized Alfv\'en waves. Consequently, it is confirmed that the results qualitatively agree with those in the axisymmetric coordinate system. That means, even in the local-spherically symmetric coordinate system, there is an upper limit on the transmitted Poynting flux into the corona when the merging height is higher ($\overline{B}=4$ G). The wind's mass loss rate and average spicule height become independent of the velocity amplitude on the photosphere ($v_{\rm conv}$) when $v_{\rm conv}/c_{s\rm ph}\gtrsim4.2\sqrt{\overline{B}/B_{\rm ph}}$ (section \ref{sec:Wave_Nonlinearity_in_the_Chromosphere}). Figure \ref{fig:fig19.eps} shows the mass loss rates of the solar wind as a function of the energy input from the photosphere. The filled and open circles are the same as those in Figure \ref{fig:fig04.eps} and the square symbols are overplotted as the results in the local-spherically symmetric coordinate system. The wind's mass loss rate in the local-spherically symmetric coordinate system with $\overline{B}=4$ G (open squares) appear to be constant for $F_{A0}\gtrsim10^{10}$ erg cm$^{-2}$ s$^{-1}$. The upper limit of mass loss rate simulated in the local-spherically symmetric coordinate system is, however, much higher than that in the axisymmetric coordinate system. This is partly because the circularly polarized Alfv\'en waves transfer the magnetic energy twice as much as the linearly polarized Alfv\'en waves when their amplitudes are the same. In other words, the critical Poynting flux $F_{A,\rm cr}$ (Eq. \ref{eq:facr_ax}) in the local-spherically symmetric coordinate system is calculated as $F_{A,\rm cr}=2\times\rho_{\rm ph}v_{\rm conv,cr}^2V_{A\rm ph}$,
\begin{equation}
  F_{A,\rm cr}=4.8\times10^{12}\ \mbox{erg cm$^{-2}$ s$^{-1}$} (\overline{B}/B_{\rm ph})
\end{equation}
As a result, the upper limit of the transmitted Poynting flux into the corona is much larger than that in the axisymmetric coordinate system. As such, the resultant mass loss rate can reach the observed level. Therefore, the abovementioned inconsistency between the observed and simulated mass loss rates in the axisymmetric coordinate system is merely an intrinsic problem of our 1D approximation.

\citet{2013PASJ...65...98S} reported their simulation results of solar and stellar winds which showed that the wind's mass loss rate saturates due to the enhanced radiative loss in the corona. From the time-steady energy equation, they paid attention to the following energy conservation law (in the local-spherically symmetric coordinate system):
\begin{align}
  &\left.\left[\rho v_rA\left({v_\perp^2\over2}+{B_\perp^2\over4\pi\rho}\right)-B_rA{v_\perp B_\perp\over4\pi}\right]\right|_{r_{\rm tc}}\nonumber\\
  &\ \ \ \approx \left.\left[\rho v_rA{v_r^2\over2}\right]\right|_\infty
  +\int^\infty_{r_{\rm tc}}AQ_{\rm rad}dr
  +\left.\left[\rho v_rA{GM_\odot\over r}\right]\right|_{r_{\rm tc}}
  \label{eq:eq_energyconserv}
\end{align}
where $v_\perp$ and $B_\perp$ are the transverse components of the velocity and magnetic field. $r_{\rm tc}$ represents the top of the chromosphere, the position with the temperature $T=2\times10^4$ K, according to the definition by \citet{2013PASJ...65...98S}. The foregoing expression means that the Poynting flux at $r=r_{\rm tc}$ (left-hand side) is converted to the kinetic energy of wind (the first term in the right-hand side) as well as the radiative loss and gravitational potential energy (the second and third terms in the right-hand side, respectively). While the kinetic energy of wind is positively correlated to the Alfv\'en waves energy at the top of the chromosphere, the large energy transmission into the corona can make the radiative energy loss dominant over the kinetic energy term. That leads to the saturation of the wind's mass loss rate. This kind of saturation is also seen in our simulation. Figure \ref{fig:fig20.eps} is analogous to Figure 8 in \citet{2013PASJ...65...98S}. It presents the comparison between the left-hand side of Eq. \ref{eq:eq_energyconserv} ($(L_Af)_{\rm tc}$) with the first and second terms in the right-hand side ($L_{\rm K,out}$ and $(L_Rf)_{\rm tc}$). As seen in this figure, $(L_Af)_{\rm tc}$ larger than $\sim4\times10^{28}$ erg s$^{-1}$ leads to the saturation of $L_{\rm K,out}$, which is associated with the enhanced $(L_Rf)_{\rm tc}$. The saturation level of $L_{\rm K,out}$ is almost consistent with that suggested by \citet{2013PASJ...65...98S} as shown below, indicated by the blue horizontal line in Figure \ref{fig:fig20.eps}.
\begin{align}
  L_{\rm K,out,sat}=2.05\times10^{28}\ \mbox{erg s}^{-1}\ (B_{\rm ph}f_{\rm ph})^{1.84}
 \label{eq:saturation_level}
\end{align}
This saturation is, however, not expected in the case of the higher merging height ($\overline{B}=4$ G). That is because the transmission of the Alfv\'en wave energy itself is limited due to its high nonlinearity in the chromosphere, as discussed in the previous subsection. This is why the open circles and squares are absent above a certain level of $(L_Af)_{\rm tc}$, indicated by the vertical thick lines in Figure \ref{fig:fig20.eps}.

\subsubsection{Spicule}

The magnetic field configuration in the spicule has been investigated using spectropolarimetric observations \citep{2005ApJ...619L.191T,2005A&A...436..325L,2015ApJ...803L..18O} or inferred from MHD seismology \citep{2007A&A...474..627Z,2008JKAS...41..173K}. However, the statistics relating the spicule dynamics to magnetic field configuration have not yet been established (see \citet{2012SSRv..169..181T} for a review). Several observational studies suggest that the different magnetic field configurations between the quiet region and coronal hole are responsible for the difference in their spicule properties, such as their height as well as ascending and transverse speeds \citep{1996ApJ...471..510J,2012ApJ...750...16Z,2012ApJ...759...18P}. Our particular attention to this relationship would be examined by future observations.

From the theoretical point of view, \citet{2016PhDT.........5I} found that the average magnetic field strength is not primarily important for the length scale of a chromospheric jet based on his 2D radiation MHD simulation. On the other hand, he noted that the scale of chromospheric jets driven by torsional motion of a flux tube is possibly dependent on the average magnetic field strength. \citet{2001ApJ...554.1151S} show that a taller spicule is associated with a lower density or stronger magnetic field based on their 1D MHD simulation. They explained, by referring to \citet{1982SoPh...78..333S}, that a taller spicule is launched by the slow shock that grows with decreasing density or less expanding flux tube. In our simulation, the average spicule height is determined by the strength of slow shock reaching the transition layer and the density scale height in the chromosphere. When $\overline{B}=29$ G, the slow shock can grow with height (Figure \ref{fig:fig13.eps}) and drive the faster spicule. The larger $v_{\rm conv}$ leads to the amplified centrifugal force and the enhanced slow shock heating, both of which could contribute to the extension of density scale height in the chromosphere (Appendix \ref{sec:density_scale_height}). As a result, the average spicule height is taller with larger $v_{\rm conv}$ in the case of $\overline{B}=29$ G (Figure \ref{fig:fig06.eps}). On the other hand, when $\overline{B}=4$ G, the intermediate shock restricts the centrifugal force from being amplified and the slow shock becomes evanescent in the upper chromosphere. This is why the average spicule height is less dependent on $v_{\rm conv}$ in the weaker $\overline{B}$.

\subsection{Limitations to Our Model and Future Perspectives}

As for the chromospheric intermediate shock, \citet{2019A&A...626A..46S} found that the decoupling of the neutral fluid against plasma can cause the intermediate shock when reconnection occurs in the partially ionized plasma. Our study as well as their study suggest that the chromospheric intermediate shock would be observed ubiquitously over the wide range of spatial scales in near future. The effect of partially ionized plasma can appear especially for the propagation of the high-frequency Alfv\'en waves \citep{2019ApJ...871....3S} and should be considered in future studies. Several limitations should be imposed on the application of the results of our study with regard to the real solar atmosphere and wind. The present study is based on a 1D approximation (symmetry assumption), flux tube model, and simplified radiation. Our study neglects the solar rotation, collisionless effects, and various wave dissipation mechanisms, including phase-mixing and turbulent dissipation \citep{2007ApJS..171..520C,2018ApJ...853..190S}. 
Nevertheless, it is worthwhile to emphasize that too highly nonlinear Alfv\'en waves in the chromosphere could restrict the energy transfer from the photosphere to the corona. As such, our findings highlight the importance of the magnetic field configuration in the chromosphere in terms of the diversity of both solar and stellar atmosphere and wind structures.

\acknowledgments
We thank K. Ichimoto for many valuable and critical comments. T.S. was supported by JSPS KAKENHI Grant Number JP18J12677. A part of this study was carried out by using the computational resources of the Center for Integrated Data Science, Institute for Space-Earth Environmental Research, Nagoya University through the joint research program, XC40 at YITP in Kyoto University, and Cray XC50 at Center for Computational Astrophysics, National Astronomical Observatory of Japan. Numerical analyses were partly carried out on analysis servers at the Center for Computational Astrophysics, National Astronomical Observatory of Japan.

\appendix
\section{Mearsurement of Mach number of Fast / Slow Shocks}
\label{sec:Mach_number_of_MHD_shock_wave}
The Alfv\'en Mach numbers of fast and slow shocks in our simulation were calculated by Eq. \ref{eq:mach_number}. The derivation is described here. Noting the subscripts $u$ and $d$ for the physical quantities in upper and downstream of the shock wave, the jump condition of momentum flux across the shock front is expressed as follows:
\begin{equation}
  \left[\rho v_x^2+p+{B_\phi^2\over 8\pi}\right]^u_d=0
\end{equation}
Since $\rho_u v_{xu}=\rho_dv_{xd}$ from the mass conservation, $\rho_uv_{xu}(v_{xu}-v_{xd})=-(p_{{\rm tot} u}-p_{{\rm tot} d})$, and thus, we have:
\begin{equation}
  M_{Au}={v_{xu}\over V_{Au}}=-{p_{{\rm tot} u}-p_{{\rm tot} d}\over\rho_uV_{Au}(v_{xu}-v_{xd})}
\end{equation}
where $p_{\rm tot}=p+B_\phi^2/(8\pi)$. The above expression is, meanwhile, not practical for the estimation of Mach number especially in the stratified atmosphere. Actually, when $v_{xu}-v_{xd}$ approaches 0, $|p_{{\rm tot} u}-p_{{\rm tot} d}|$ tends to $\rho g_\odot\Delta x$ ($\Delta x$ is the discretization) because of the stratification. This leads to the overestimation of the Mach number for weak shock in the lower atmosphere. In order to correct it, we used the following formula:
\begin{equation}
  M_{Au}=-{1\over V_A}\left({\partial v_x\over\partial x}\right)^{-1}\left\{{1\over\rho}{\partial\over\partial x}\left(p+{B_\phi^2\over8\pi}\right)-{\partial\over\partial x}\left({GM_\odot\over r}\right)\right\}
\end{equation}

\section{Density Scale Height of the Atmosphere}
\label{sec:density_scale_height}
\begin{figure}
  \begin{center}
    \epsscale{1.2}
    \plotone{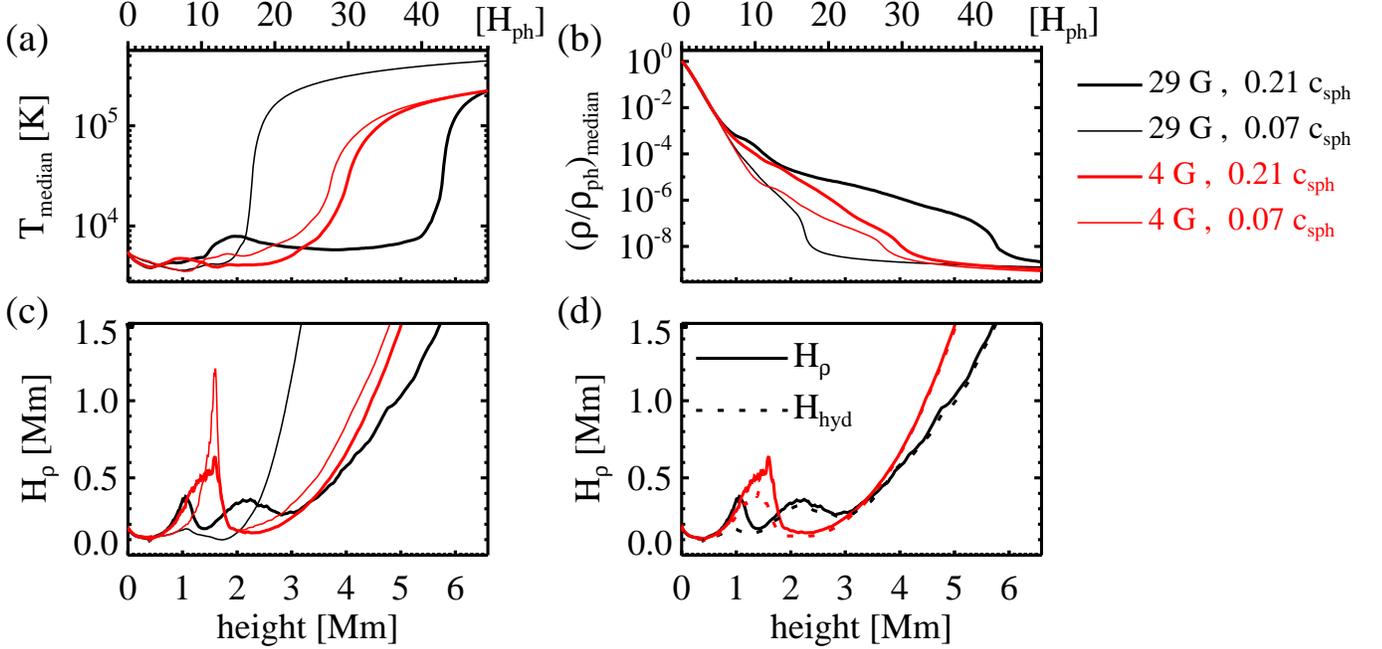}
    \caption{
      The median profile of (a) temperature, (b) density, and the averaged density scale height (c) in lower atmosphere. The black and red lines correspond to the simulation results in the cases of $\overline{B}=29$ and 4 G, respectively. The thin and thick lines represent the results with $v_{\rm conv}/c_{s\rm ph}=0.07$ and 0.21, respectively. In panel (d), the density scale heights in the case of $v_{\rm conv}/c_{s\rm ph}=0.21$ are compared to their constituents related to the stratification by the gravitational acceleration and temperature gradient (dotted lines).
    }
    \label{fig:fig21.eps}
  \end{center}
\end{figure}
The density scale height of the atmosphere follows the dynamic equilibrium determined by the poloidal component of the equation of motion.
\begin{equation}
  {\partial v_x\over\partial t}+v_x{\partial v_x\over\partial x}
  +{1\over\rho}{\partial p\over\partial x}
  +{1\over\rho}{\partial\over\partial x}\left(B_\phi^2\over8\pi\right)
  +{B_\phi^2\over4\pi\rho}{\partial\ln\sqrt{A}\over\partial x}
  -v_\phi^2{\partial\ln\sqrt{A}\over\partial x}-{\partial\over\partial x}\left(GM_\odot\over r\right)=0
\end{equation}
By substituting $p=\rho\times(p/\rho)$ and considering the temporal average, the following is obtained.
\begin{equation}
  -\left\langle{\partial\ln\rho\over\partial x}\right\rangle
  =\left\langle{\rho v_x\over p}{\partial v_x\over\partial x}\right\rangle
  +\left\langle{1\over p}{\partial\over\partial x}\left(B_\phi^2\over8\pi\right)\right\rangle
  +\left\langle{B_\phi^2\over4\pi p}{\partial\ln\sqrt{A}\over\partial x}\right\rangle
  -\left\langle{\rho v_\phi^2\over p}{\partial\ln\sqrt{A}\over\partial x}\right\rangle
  +\left\langle{\rho\over p}{\partial\over\partial x}\left({p\over\rho}-{GM_\odot\over r}\right)\right\rangle
\end{equation}
The left-hand side represents the reciprocal of the density scale height and is expressed with the harmonic mean of the several scale heights.
\begin{equation}
  {1\over H_\rho}={1\over H_{\rm dyn}}+{1\over H_{\rm Bp}}+{1\over H_{\rm Bt}}+{1\over H_{\rm cnt}}+{1\over H_{\rm hyd}}
\end{equation}
Here, $H_{\rm dyn}$, $H_{\rm Bp}$, $H_{\rm Bt}$, and $H_{\rm cnt}$ represent the scale heights which are related to the dynamic pressure, magnetic pressure, magnetic tension force, and centrifugal force, respectively.
\begin{equation}
  {1\over H_\rho}=-\left\langle{\partial\ln\rho\over\partial x}\right\rangle, \ \ \ \ \ 
  {1\over H_{\rm dyn}}=\left\langle{\rho\over p}v_x{\partial v_x\over\partial x}\right\rangle,
\end{equation}
\begin{equation}
  {1\over H_{\rm Bp}}=\left\langle{1\over p}{\partial\over\partial x}\left(B_\phi^2\over8\pi\right)\right\rangle, \ \ \ \ \
  {1\over H_{\rm Bt}}=\left\langle{B_\phi^2\over4\pi p}{\partial\ln\sqrt{A}\over\partial x}\right\rangle,
\end{equation}
\begin{equation}
  {1\over H_{\rm cnt}}=-\left\langle{\rho v_\phi^2\over p}{\partial\ln\sqrt{A}\over\partial x}\right\rangle,
  \ \ \ \ \ 
  {1\over H_{\rm hyd}}=\left\langle{\rho\over p}{\partial\over\partial x}\left({p\over\rho}-{GM_\odot\over r}\right)\right\rangle
\end{equation}
$H_{\rm hyd}$ is the density scale height of the atmosphere in the hydrostatic equilibrium when the temperature profile is given. For the isothermal atmosphere, $H_{\rm hyd}$ is expressed as follows:
\begin{equation}
  {1\over H_{\rho}}={1\over H_{\rm hyd}}={\mu g_\odot\over R_gT}{r_\odot^2\over r^2}{dr\over dx}
\end{equation}
In the expanding flux tube, $H_{\rm cnt}$ and $H_{\rm B_t}$ are always negative and positive, respectively. $H_{\rm B_p}$ and $H_{\rm dyn}$ are also usually negative and positive, respectively. These correspond to the acceleration by the magnetic pressure gradient and centrifugal force, as well as the deceleration by the magnetic tension force and dynamical pressure gradient.

Figure \ref{fig:fig21.eps} shows the dependence of the density scale height on $\overline{B}$ and $v_{\rm conv}$. 
In Figure \ref{fig:fig21.eps} (c), $H_\rho$ is plotted to see its dependence on $v_{\rm conv}$ and $\overline{B}$. The black and red lines represent the results in the $\overline{B}$=29 and 4 G cases, respectively. The thin and thick lines correspond to the results in the $v_{\rm conv}/c_{s\rm ph}=$0.07 and 0.21 cases, respectively. There are two local maxima around $\sim$ 1 Mm and 2.2 Mm in the profile of $H_\rho$ when $(\overline{B},v_{\rm conv}/c_{s\rm ph})=$(29 G, 0.21) (black thick line). Both of them are not seen in the $v_{\rm conv}/c_{s\rm ph}=0.07$ case (black thin line). On the other hand, when $\overline{B}=4$ G, the profiles of $H_\rho$ have a single maximum, regardless of the $v_{\rm conv}$. In Figure \ref{fig:fig21.eps} (d), focus is placed on the case of $v_{\rm conv}/c_{s\rm ph}=0.21$ and $H_\rho$ is compared to $H_{\rm hyd}$.
$H_{\rho}$ and $H_{\rm hyd}$ remarkably disagree with each other around 1 Mm in the case of $\overline{B}=29$ G, while they agree around 2.2 Mm. These suggest that the first local maximum of $H_{\rho}$ is accounted for by the magnetic pressure gradient and the centrifugal force while the second one results from higher chromospheric temperature.
Compared to the case of $v_{\rm conv}/c_{s\rm ph}=0.07$, when $v_{\rm conv}/c_{s\rm oh}=0.21$, Alfv\'en waves in the lower chromosphere are naturally amplified and the temperature in the upper chromosphere is increased due to the heating by the slow shock. This leads to the two local maxima in the profile of $H_\rho$.
The single local maximum in the $\overline{B}=4$ G case corresponds to the first local maximum in the case of $(\overline{B},v_{\rm conv}/c_{s\rm ph})=$(29 G, 0.21). This implies that the chromospheric density scale height does not extend even with a larger $v_{\rm conv}$ because the chromospheric temperature is less dependent on it compared to that in the case of $\overline{B}=29$ G. This less dependence of chromospheric temperature on $v_{\rm conv}$ would result from the evanescence of slow shock in the upper chromosphere with a weak magnetic field.

\section{Axisymmetric and Local-spherically Symmetric Coordinate Systems}
\label{sec:appendix_coordinate}

We note the different curvilinear coordinate systems which have been traditionally employed in 1D models. The derivation of the basic equations in each coordinate system is described here.

The most general expression of our basic equations in the curvilinear coordinate system are written as follows:
\begin{equation}
  {\partial\rho\over\partial t}+{1\over h_1h_2h_3}\sum_{\epsilon_{ijk}=1}{\partial\over\partial x_i}(h_jh_k\rho v_i)=0
\end{equation}
\begin{align}
  {\partial\over\partial t}\left({p\over\gamma-1}+{\rho v^2\over2}+{B^2\over8\pi}\right)
  +&{1\over h_1h_2h_3}\sum_{\epsilon_{ijk}=1}{\partial\over\partial x_i}\left[h_jh_k\left({\gamma p\over\gamma-1}+{\rho v^2\over2}+{B_j^2+B_k^2\over4\pi}\right)v_i
    -h_jh_k{B_jv_j+B_kv_k\over4\pi}B_i\right]\nonumber\\
  =&\sum_i\rho v_i{1\over h_i}{\partial\over\partial x_i}\left(GM_\odot\over r\right)
  -{1\over h_1h_2h_3}\sum_{\epsilon_{ijk}=1}{\partial\over\partial x_i}(h_jh_kF_{{\rm c}i})-Q_{\rm rad}
\end{align}
\begin{align}
  {\partial(\rho v_i)\over\partial t}+&{1\over h_i}{\partial\Pi_{ii}\over\partial x_i}
  +{1\over h_ih_1h_2h_3}\left\{{\partial\over\partial x_j}(h_kh_i^2\Pi_{ji})
    +{\partial\over\partial x_k}(h_i^2h_j\Pi_{ki})\right\}\nonumber\\
    =&{1\over h_i}\left\{{\Pi_{jj}-\Pi_{ii}\over h_j}{\partial h_j\over\partial x_i}
    +{\Pi_{kk}-\Pi_{ii}\over h_k}{\partial h_k\over\partial x_i}
    +\rho{\partial\over\partial x_i}\left({GM_\odot\over r}\right)\right\}
\end{align}
\begin{equation}
  {\partial B_i\over\partial t}+{1\over h_jh_k}\left[{\partial\over\partial x_j}\{h_k(v_jB_i-v_iB_j)\}+{\partial\over\partial x_k}\{h_j(v_kB_i-v_iB_k)\}\right]=0
\end{equation}
\begin{equation}
  \sum_{\epsilon_{ijk}=1}{\partial\over\partial x_i}(h_jh_kB_i)=0
\end{equation}
where $h_1$, $h_2$, $h_3$ are the scale factors of the curvilinear coordinate system. $v^2=v_1^2+v_2^2+v_3^2$, $B^2=B_1^2+B_2^2+B_3^2$, $\Pi_{ij}=\{p+B^2/(8\pi)\}\delta_{ij}+\rho v_iv_j-B_iB_j/(4\pi)$. $\sum_{\epsilon_{ijk}=1}$ means the summation over a set of even permutation of $(1,2,3)$.

\begin{figure*}
  \begin{center}
    \epsscale{.8}
    \plotone{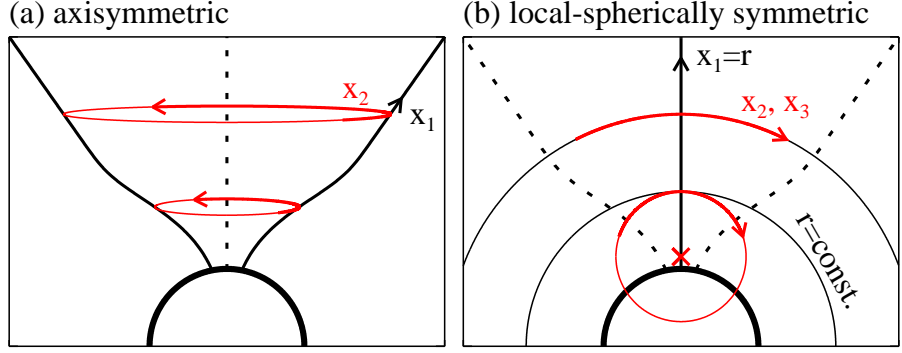}
    \caption{The difference between the (a) axisymmetric and (b) local-spherically symmetric coordinate systems. The poloidal axis $x_1$ is represented by solid black arrows and the toroidal or transverse axis $x_2$ is represented by thick red arrows. The poloidal axis $x_1$ of the local-spherically symmetric coordinate system agrees with the radial axis, while that of the axisymmetric coordinate system agrees with that of the flux tube. The example of “local sphere” in the local-spherically symmetric coordinate system is shown with the thin red circle in panel (b), the center of which corresponds to the red $\times$ symbol. The local sphere is not the same as the sphere with $r=$ const. unless the magnetic flux tube expands radially (i.e., $B_1\propto r^{-2}$).}
    \label{fig:fig22.eps}
  \end{center}
\end{figure*}

There are two traditional approaches in simplifying the abovementioned equations into the 1D configuration. The first one is the axisymmetric coordinate system based on the assumption that $\partial_2=0$ for $h_2$, $h_3$, $r$, other physical quantities, and that $B_3=0$ and $v_3=0$ \citep{1982SoPh...75...35H,1999ApJ...514..493K,2010ApJ...710.1857M}. The poloidal axis $x_1$ represents the outer edge of the magnetic flux tube (Figure \ref{fig:fig22.eps}). The second is the local-spherically symmetric coordinate system based on the assumption that $\partial_2=0$ and $\partial_3=0$ for $h_2$, $h_3$, $r$, and other physical quantities \citep{2005ApJ...632L..49S,2006JGRA..111.6101S,2018ApJ...854....9S}. The poloidal axis $x_1$ in this case agrees with the radial axis of the spherical coordinate system. The scale factors $h_2$ and $h_3$ are specified so that $h_2h_3\propto B_1^{-1}$ is along the $x_1$-axis in both coordinate systems. Thereafter, they can be expressed as $h_{2,3}^{-1}=|\partial_1(\ln\sqrt{B_1})|$, which is close to $h_{2,3}=r$ in the distance where $B_1\propto r^{-2}$. By noting $x_1$, $x_2$, and $x_3$ with $x$, $\phi$, and $x_3$ for the axisymmetric coordinate system or with $x(=r)$, $y$, and $z$ for the local-spherically symmetric coordinate system, the following equation systems are obtained:

In the axisymmetric coordinate system,
\begin{equation}
  {\partial\rho\over\partial t}+{1\over A}{\partial\over\partial x}(\rho v_xA)=0
\end{equation}
\begin{align}
  {\partial\over\partial t}\left({p\over\gamma-1}+{1\over2}\rho v^2+{B^2\over8\pi}\right)
  +&{1\over A}{\partial\over\partial x}\left[A\left\{\left({\gamma p\over\gamma-1}+{\rho v^2\over2}+{B_\phi^2\over4\pi}\right)v_x-{B_x\over4\pi}(B_\phi v_\phi)\right\}\right]\nonumber\\
  &=\rho v_x{\partial\over\partial x}\left({GM_\odot\over r}\right)-{1\over A}{\partial\over\partial x}(AF_{\rm c})-Q_{\mbox{\scriptsize rad}}
\end{align}
\begin{equation}
  {\partial(\rho v_x)\over\partial t}+{\partial p\over\partial x}+{1\over A}{\partial\over\partial x}\left\{\left(\rho v_x^2+{B_\phi^2\over8\pi}\right)A\right\}
  -\rho v_\phi^2{\partial\ln\sqrt{A}\over\partial x}-\rho{\partial\over\partial x}\left({GM_\odot\over r}\right)=0
\end{equation}
\begin{equation}
  {\partial(\rho v_\phi)\over\partial t}+{1\over A\sqrt{A}}
  {\partial\over\partial x}\left\{A\sqrt{A}\left(\rho v_xv_\phi-{B_xB_\phi\over4\pi}\right)\right\}=0
\end{equation}
\begin{equation}
  {\partial B_\phi\over\partial t}+{1\over \sqrt{A}}{\partial\over\partial x}\Big(\sqrt{A}(v_xB_\phi-v_\phi B_x)\Big)=0
\end{equation}
\begin{equation}
  B_xA=\mbox{const.}
\end{equation}
\begin{equation}
  {dx\over dr}=\sqrt{1+\left({d\sqrt{A}\over dr}\right)^2}
\end{equation}

In the local-spherically symmetric coordinate system,
\begin{equation}
  {\partial\rho\over\partial t}+{1\over A}{\partial\over\partial x}(\rho v_xA)=0\label{eq:mass-spherically}
\end{equation}
\begin{align}
  {\partial\over\partial t}\left({p\over\gamma-1}+{1\over2}\rho v^2+{B^2\over8\pi}\right)
  +&{1\over A}{\partial\over\partial x}\left[A\left\{\left({\gamma p\over\gamma-1}+{\rho v^2\over2}+{B_\perp^2\over4\pi}\right)v_x-{B_x\bm B_\perp\cdot\bm v_\perp\over4\pi}\right\}\right]\nonumber\\
  &=-\rho v_x{GM_\sun\over r^2}-{1\over A}{\partial\over\partial x}(AF_{\rm c})-Q_{\rm rad}
\end{align}
\begin{equation}
  {\partial(\rho v_x)\over\partial t}+{\partial p\over\partial x}+{1\over A}{\partial\over\partial x}\left\{\left(\rho v_x^2+{B_\perp^2\over8\pi}\right)A\right\}-\rho v_\perp^2{\partial\ln\sqrt{A}\over\partial x}+\rho{GM_\sun\over r^2}=0
\end{equation}
\begin{equation}
  {\partial(\rho\bm v_\perp)\over\partial t}+{1\over A\sqrt{A}}
  {\partial\over\partial x}\left\{A\sqrt{A}\left(\rho v_x\bm v_\perp-{B_x\bm B_\perp\over4\pi}\right)\right\}=0
\end{equation}
\begin{equation}
  {\partial\bm B_\perp\over\partial t}+{1\over \sqrt{A}}{\partial\over\partial x}\Big(\sqrt{A}(v_x\bm B_\perp-\bm v_\perp B_x)\Big)=0
\end{equation}
\begin{equation}
  B_xA=\mbox{const.}
\end{equation}
where $\bm B_\perp=(B_y,B_z)$ and $\bm v_\perp=(v_y,v_z)$. Since the two transverse components of the velocity and magnetic field are taken into account in the local-spherically symmetric coordinate system, the circularly polarized Alfv\'en waves can be discussed only by using this coordinate system. However, it should be noted that the simulation result based on the local-spherically symmetric coordinate system is not always representative of the dynamics of the magnetic flux tube in the 3D space, especially for the low frequency Alfv\'en waves in the lower atmosphere wherein the flux tube expands super radially and the gravity cannot be neglected. This is because it is not always possible to assume both that $\partial_x(h_yh_zB_x)=0$ and that $\partial_yr=\partial_zr=0$, as required in the local-spherically symmetric coordinate system. In fact, because the local sphere with the curvature radius of $h_y$ is not identical to the sphere radius of $r$ unless $B_x\propto r^{-2}$, the gravitational acceleration is not uniform on the $yz$ plane. Therefore, when $B_x$ expands more strongly than $r^{-2}$, we can assume $\partial_yr=\partial_zr=0$ only along the specific direction where the $x$-axis agrees with the radial axis, and the gravity term, which depends on $\partial_{y,z}r$, affects the transverse components of the equation of motion anywhere else. This is why the assumptions that $B_3=0$ and $v_3=0$ are imposed in the axisymmetric coordinate system. The magnitude of the gravitational acceleration in the $y$-component of the equation of motion around $(y,z)=(0,0)$ is estimated as $\rho g_\odot y/R$, where $R=|\partial_x\ln\sqrt{B_x}|^{-1}$ is the curvature radius of the $y$-axis. When the flux tube expands exponentially with the pressure scale height $H_p$ in the lower atmosphere, we find $B_x\propto e^{-(r-r_\odot)/(2H_p)}$ and, thus, $R\sim 4H_p$ (see section \ref{sec:magnetic_flux_tube_model}). For the propagation of the Alfv\'en waves with the wavelength $\lambda_A$ and the frequency $\nu_A$, this gravitational acceleration is not negligible compared to the restoring force $\sim B_xB_y/(4\pi\lambda_A)$. In fact, by using $y\sim v_y/\nu_A$ and $B_y/v_y\sim\sqrt{4\pi\rho}$, it is obtained that $[\rho g_\odot v_y/\nu_A/(4H_p)]/[B_xB_y/(4\pi\lambda_A)]\sim\nu_{\rm ac}^2/\nu_A^2$, where $\nu_{\rm ac}$ is the acoustic cut-off frequency. This means that the assumption of local-spherically symmetric coordinate system is not appropriate in describing the propagation of the Alfv\'en waves with a frequency lower than the acoustic cut-off frequency in the 3D space.




\begin{thebibliography}{}
\bibitem[Alfv{\'e}n(1947)]{1947MNRAS.107..211A} Alfv{\'e}n, H.\ 1947, \mnras, 107, 211
\bibitem[An et al.(1990)]{1990ApJ...350..309A} An, C.-H., Suess, S.~T., Moore, R.~L., \& Musielak, Z.~E.\ 1990, \apj, 350, 309
\bibitem[Antolin \& Shibata(2010)]{2010ApJ...712..494A} Antolin, P., \& Shibata, K.\ 2010, \apj, 712, 494
\bibitem[Bale et al.(2013)]{2013ApJ...769L..22B} Bale, S.~D., Pulupa, M., Salem, C., Chen, C.~H.~K., \& Quataert, E.\ 2013, \apjl, 769, L22
\bibitem[Banerjee et al.(2009)]{2009A&A...501L..15B} Banerjee, D., P{\'e}rez-Su{\'a}rez, D., \& Doyle, J.~G.\ 2009, \aap, 501, L15
\bibitem[Bavassano et al.(1982)]{1982JGR....87.3617B} Bavassano, B., Dobrowolny, M., Mariani, F., et al.\ 1982, \jgr, 87, 3617
\bibitem[Bavassano et al.(2001)]{2001JGR...10610659B} Bavassano, B., Pietropaolo, E., \& Bruno, R.\ 2001, \jgr, 106, 10659
\bibitem[Belcher \& Davis(1971)]{1971JGR....76.3534B} Belcher, J.~W., \& Davis, L.\ 1971, \jgr, 76, 3534
\bibitem[Belcher \& MacGregor(1976)]{1976ApJ...210..498B} Belcher, J.~W., \& MacGregor, K.~B.\ 1976, \apj, 210, 498
\bibitem[Carlsson \& Leenaarts(2012)]{2012A&A...539A..39C} Carlsson, M., \& Leenaarts, J.\ 2012, \aap, 539, A39
\bibitem[Cirtain et al.(2007)]{2007Sci...318.1580C} Cirtain, J.~W., Golub, L., Lundquist, L., et al.\ 2007, Science, 318, 1580
\bibitem[Coleman(1968)]{1968ApJ...153..371C} Coleman, P.~J.\ 1968, \apj, 153, 371
\bibitem[Cranmer \& van Ballegooijen(2005)]{2005ApJS..156..265C} Cranmer, S.~R., \& van Ballegooijen, A.~A.\ 2005, \apjs, 156, 26
\bibitem[Cranmer et al.(2007)]{2007ApJS..171..520C} Cranmer, S.~R., van Ballegooijen, A.~A., \& Edgar, R.~J.\ 2007, \apjs, 171, 520
\bibitem[De Pontieu et al.(2007)]{2007Sci...318.1574D} De Pontieu, B., McIntosh, S.~W., Carlsson, M., et al.\ 2007, Science, 318, 1574
\bibitem[Derby(1978)]{1978ApJ...224.1013D} Derby, N.~F., Jr.\ 1978, \apj, 224, 1013
\bibitem[Goldstein(1978)]{1978ApJ...219..700G} Goldstein, M.~L.\ 1978, \apj, 219, 700 
\bibitem[Hahn \& Savin(2013)]{2013ApJ...776...78H} Hahn, M., \& Savin, D.~W.\ 2013, \apj, 776, 78
\bibitem[Harvey et al.(1982)]{1982SoPh...79..149H} Harvey, K.~L., Sheeley, N.~R., \& Harvey, J.~W.\ 1982, \solphys, 79, 149
\bibitem[Hasan et al.(2003)]{2003ApJ...585.1138H} Hasan, S.~S., Kalkofen, W., van Ballegooijen, A.~A., \& Ulmschneider, P.\ 2003, \apj, 585, 1138
\bibitem[Hasan et al.(2005)]{2005ApJ...631.1270H} Hasan, S.~S., van Ballegooijen, A.~A., Kalkofen, W., \& Steiner, O.\ 2005, \apj, 631, 1270 
\bibitem[Heggland et al.(2011)]{2011ApJ...743..142H} Heggland, L., Hansteen, V.~H., De Pontieu, B., et al.\ 2011, \apj, 743, 142
\bibitem[Heinemann \& Olbert(1980)]{1980JGR....85.1311H} Heinemann, M., \& Olbert, S.\ 1980, \jgr, 85, 1311 
\bibitem[Heyvaerts \& Priest(1983)]{1983A&A...117..220H} Heyvaerts, J., \& Priest, E.~R.\ 1983, \aap, 117, 220
\bibitem[Hollweg(1971)]{1971JGR....76.5155H} Hollweg, J.~V.\ 1971, \jgr, 76, 5155
\bibitem[Hollweg(1978)]{1978SoPh...56..305H} Hollweg, J.~V.\ 1978, \solphys, 56, 305
\bibitem[Hollweg et al.(1982)]{1982SoPh...75...35H} Hollweg, J.~V., Jackson, S., \& Galloway, D.\ 1982, \solphys, 75, 35
\bibitem[Hollweg(1992)]{1992ApJ...389..731H} Hollweg, J.~V.\ 1992, \apj, 389, 731
\bibitem[Hori et al.(1997)]{1997ApJ...489..426H} Hori, K., Yokoyama, T., Kosugi, T., et al.\ 1997, \apj, 489, 426
\bibitem[Iijima \& Yokoyama(2015)]{2015ApJ...812L..30I} Iijima, H., \& Yokoyama, T.\ 2015, \apjl, 812, L30
\bibitem[Iijima(2016)]{2016PhDT.........5I} Iijima, H.\ 2016, Ph.D. Thesis
\bibitem[Johannesson, \& Zirin(1996)]{1996ApJ...471..510J} Johannesson, A., \& Zirin, H.\ 1996, \apj, 471, 510
\bibitem[Kim et al.(2008)]{2008JKAS...41..173K} Kim, Y.-H., Bong, S.-C., Park, Y.-D., et al.\ 2008, Journal of Korean Astronomical Society, 41, 173
\bibitem[Kopp \& Holzer(1976)]{1976SoPh...49...43K} Kopp, R.~A., \& Holzer, T.~E.\ 1976, \solphys, 49, 43 
\bibitem[Kudoh \& Shibata(1999)]{1999ApJ...514..493K} Kudoh, T., \& Shibata, K.\ 1999, \apj, 514, 493
\bibitem[L{\'o}pez Ariste, \& Casini(2005)]{2005A&A...436..325L} L{\'o}pez Ariste, A., \& Casini, R.\ 2005, \aap, 436, 325
\bibitem[Marsch et al.(1989)]{1989JGR....94.6893M} Marsch, E., Pilipp, W.~G., Thieme, K.~M., et al.\ 1989, \jgr, 94, 6893
\bibitem[Matsumoto \& Kitai(2010)]{2010ApJ...716L..19M} Matsumoto, T., \& Kitai, R.\ 2010, \apjl, 716, L19
\bibitem[Matsumoto \& Shibata(2010)]{2010ApJ...710.1857M} Matsumoto, T., \& Shibata, K.\ 2010, \apj, 710, 1857
\bibitem[Matsumoto \& Suzuki(2012)]{2012ApJ...749....8M} Matsumoto, T., \& Suzuki, T.~K.\ 2012, \apj, 749, 8
\bibitem[Matsumoto \& Suzuki(2014)]{2014MNRAS.440..971M} Matsumoto, T., \& Suzuki, T.~K.\ 2014, \mnras, 440, 971
\bibitem[Meyer et al.(2012)]{2012MNRAS.422.2102M} Meyer, C.~D., Balsara, D.~S., \& Aslam, T.~D.\ 2012, \mnras, 422, 2102
\bibitem[Meyer et al.(2014)]{2014JCoPh.257..594M} Meyer, C.~D., Balsara, D.~S., \& Aslam, T.~D.\ 2014, Journal of Computational Physics, 257, 594
\bibitem[Miyoshi \& Kusano(2005)]{2005JCoPh.208..315M} Miyoshi, T., \& Kusano, K.\ 2005, Journal of Computational Physics, 208, 315
\bibitem[Moriyasu et al.(2004)]{2004ApJ...601L.107M} Moriyasu, S., Kudoh, T., Yokoyama, T., et al.\ 2004, \apjl, 601, L107
\bibitem[Nagai(1980)]{1980SoPh...68..351N} Nagai, F.\ 1980, \solphys, 68, 351
\bibitem[Okamoto \& De Pontieu(2011)]{2011ApJ...736L..24O} Okamoto, T.~J., \& De Pontieu, B.\ 2011, \apjl, 736, L24
\bibitem[Orozco Su{\'a}rez et al.(2015)]{2015ApJ...803L..18O} Orozco Su{\'a}rez, D., Asensio Ramos, A., \& Trujillo Bueno, J.\ 2015, \apjl, 803, L18
\bibitem[Osterbrock(1961)]{1961ApJ...134..347O} Osterbrock, D.~E.\ 1961, \apj, 134, 347
\bibitem[Parker(1964)]{1964ApJ...139...93P} Parker, E.~N.\ 1964, \apj, 139, 93
\bibitem[Pereira et al.(2012)]{2012ApJ...759...18P} Pereira, T.~M.~D., De Pontieu, B., \& Carlsson, M.\ 2012, \apj, 759, 18
\bibitem[Saito et al.(2001)]{2001ApJ...554.1151S} Saito, T., Kudoh, T., \& Shibata, K.\ 2001, \apj, 554, 1151
\bibitem[Shibata \& Suematsu(1982)]{1982SoPh...78..333S} Shibata, K., \& Suematsu, Y.\ 1982, \solphys, 78, 333
\bibitem[Shibata \& Uchida(1985)]{1985PASJ...37...31S} Shibata, K., \& Uchida, Y.\ 1985, \pasj, 37, 31
\bibitem[Shoda et al.(2018)]{2018ApJ...853..190S} Shoda, M., Yokoyama, T., \& Suzuki, T.~K.\ 2018, \apj, 853, 190
\bibitem[Shoda \& Yokoyama(2018)]{2018ApJ...854....9S} Shoda, M., \& Yokoyama, T.\ 2018, \apj, 854, 9
\bibitem[Shoda et al.(2019)]{2019ApJ...880L...2S} Shoda, M., Suzuki, T.~K., Asgari-Targhi, M., et al.\ 2019, \apjl, 880, L2
\bibitem[Shu \& Osher(1988)]{1988JCoPh..77..439S} Shu, C.-W., \& Osher, S.\ 1988, Journal of Computational Physics, 77, 439
\bibitem[Snow \& Hillier(2019)]{2019A&A...626A..46S} Snow, B., \& Hillier, A.\ 2019, \aap, 626, A46 
\bibitem[Soler et al.(2019)]{2019ApJ...871....3S} Soler, R., Terradas, J., Oliver, R., \& Ballester, J.~L.\ 2019, \apj, 871, 3 
\bibitem[Spitzer \& H{\"a}rm(1953)]{1953PhRv...89..977S} Spitzer, L. \& H{\"a}rm, R.\ 1953, Physical Review, 89, 977
\bibitem[Suzuki \& Inutsuka(2005)]{2005ApJ...632L..49S} Suzuki, T.~K., \& Inutsuka, S.-i.\ 2005, \apjl, 632, L49
\bibitem[Suzuki \& Inutsuka(2006)]{2006JGRA..111.6101S} Suzuki, T.~K., \& Inutsuka, S.-I.\ 2006, Journal of Geophysical Research (Space Physics), 111, A06101 
\bibitem[Suzuki(2006)]{2006ApJ...640L..75S} Suzuki, T.~K.\ 2006, \apjl, 640, L75
\bibitem[Suzuki(2007)]{2007ApJ...659.1592S} Suzuki, T.~K.\ 2007, \apj, 659, 1592
\bibitem[Suzuki et al.(2013)]{2013PASJ...65...98S} Suzuki, T.~K., Imada, S., Kataoka, R., et al.\ 2013, \pasj, 65, 98
\bibitem[Suzuki(2018)]{2018PASJ...70...34S} Suzuki, T.~K.\ 2018, \pasj, 70, 34
\bibitem[Terasawa et al.(1986)]{1986JGR....91.4171T} Terasawa, T., Hoshino, M., Sakai, J.-I., \& Hada, T.\ 1986, \jgr, 91, 4171 
\bibitem[Trujillo Bueno et al.(2005)]{2005ApJ...619L.191T} Trujillo Bueno, J., Merenda, L., Centeno, R., et al.\ 2005, \apjl, 619, L191
\bibitem[Tsiropoula et al.(2012)]{2012SSRv..169..181T} Tsiropoula, G., Tziotziou, K., Kontogiannis, I., et al.\ 2012, \ssr, 169, 181
\bibitem[Velli(1993)]{1993A&A...270..304V} Velli, M.\ 1993, \aap, 270, 304
\bibitem[Wang et al.(2000)]{2000GeoRL..27..505W} Wang, Y.-M., Lean, J., \& Sheeley, N.~R., Jr.\ 2000, \grl, 27, 505 
\bibitem[Wang(2010)]{2010ApJ...715L.121W} Wang, Y.-M.\ 2010, \apjl, 715, L121 
\bibitem[Wang \& Yokoyama(2020)]{2020ApJ...891..110W} Wang, Y. \& Yokoyama, T.\ 2020, \apj, 891, 110
\bibitem[Washinoue \& Suzuki(2019)]{2019ApJ...885..164W} Washinoue, H., \& Suzuki, T.~K.\ 2019, \apj, 885, 164
\bibitem[Wiegelmann \& Solanki(2004)]{2004SoPh..225..227W} Wiegelmann, T., \& Solanki, S.~K.\ 2004, \solphys, 225, 227
\bibitem[Withbroe, \& Noyes(1977)]{1977ARA&A..15..363W} Withbroe, G.~L., \& Noyes, R.~W.\ 1977, \araa, 15, 363
\bibitem[Withbroe(1989)]{1989ApJ...337L..49W} Withbroe, G.~L.\ 1989, \apjl, 337, L49
\bibitem[Yasuda et al.(2019)]{2019ApJ...879...77Y} Yasuda, Y., Suzuki, T.~K., \& Kozasa, T.\ 2019, \apj, 879, 77
\bibitem[Zaqarashvili et al.(2007)]{2007A&A...474..627Z} Zaqarashvili, T.~V., Khutsishvili, E., Kukhianidze, V., et al.\ 2007, \aap, 474, 627
\bibitem[Zhang et al.(2012)]{2012ApJ...750...16Z} Zhang, Y.~Z., Shibata, K., Wang, J.~X., et al.\ 2012, \apj, 750, 16
\end{thebibliography}
\end{document}